\documentclass[num-refs]{wiley-article}
\usepackage{color}
\usepackage{ulem}

\usepackage{mathtools}
\usepackage{booktabs}
\usepackage[english]{babel} % English language/hyphenation
\usepackage{amsfonts}
\usepackage{enumitem}
\usepackage{graphicx}
\usepackage[toc,page]{appendix}

\usepackage{booktabs} % Horizontal rules in tables
\usepackage{natbib}
\usepackage{setspace}
\usepackage{microtype} % Slightly tweak font spacing for aesthetics
\usepackage{tabularx}
\usepackage{subcaption}
\usepackage{caption}
\usepackage{siunitx}
\usepackage{footnote}
\usepackage{algorithm}% http://ctan.org/pkg/algorithms
\usepackage[noend]{algpseudocode}% http://ctan.org/pkg/algorithmicx
\usepackage{multirow}
\papertype{Original Article}
\paperfield{Journal Section}
\makeatletter
\newcommand{\distas}[1]{\mathbin{\overset{#1}{\kern\z@\sim}}}%
\newsavebox{\mybox}\newsavebox{\mysim}
\newcommand{\distras}[1]{%
  \savebox{\mybox}{\hbox{\kern3pt$\scriptstyle#1$\kern3pt}}%
  \savebox{\mysim}{\hbox{$\sim$}}%
  \mathbin{\overset{#1}{\kern\z@\resizebox{\wd\mybox}{\ht\mysim}{$\sim$}}}%
}
\newcolumntype{C}[1]{>{\centering\let\newline\\\arraybackslash\hspace{0pt}}m{#1}}

\title{Efficient Calibration for Imperfect Epidemic Models with Applications to the Analysis of COVID-19}

\abbrevs{Efficient Calibration for Imperfect Epidemic Models}

\author[1]{Chih-Li Sung}
\author[2]{Ying Hung}
\affil[1]{Department of Statistics and Probability, Michigan State University, East Lansing, USA}
\affil[2]{Department of Statistics, Rutgers, the State University of New Jersey, New Brunswick, USA}

\corraddress{Chih-Li Sung, Department of Statistics and Probability, Michigan State University, East Lansing, USA}
\corremail{sungchih@msu.edu}

\fundinginfo{This work was supported by NSF DMS 1660477 and NSF HDR TRIPODS award CCF 1934924}

\runningauthor{C.-L. Sung and Y. Hung}

\begin{document}

\maketitle

\begin{abstract}
The estimation of unknown parameters in simulations, also known as calibration, is crucial for practical management of epidemics and prediction of pandemic risk. A simple yet widely used approach is to estimate the parameters by minimizing the sum of the squared distances between actual observations and simulation outputs. It is shown in this paper that this method is inefficient, particularly when the epidemic models are developed based on certain simplifications of reality, also known as imperfect models which are commonly used in practice. To address this issue, a new estimator is introduced that is asymptotically consistent, has a smaller estimation variance than the least squares estimator, and achieves the semiparametric efficiency. Numerical studies are performed to examine the finite sample performance. The proposed method is applied to the analysis of the COVID-19 pandemic for 20 countries based on the SEIR (Susceptible-Exposed-Infectious-Recovered) model with both deterministic and stochastic simulations. The estimation of the parameters, including the basic reproduction number and the average incubation period, reveal the risk of disease outbreaks in each country and provide insights to the design of public health interventions.

\keywords{Compartmental models, Basic reproduction number, Stochastic simulations, Kernel Poisson regression, Semiparametric efficiency.}
\end{abstract}

%  Please place your key words in alphabetical order, separated
%  by semicolons, with the first letter of the first word capitalized,
%  and a period at the end of the list.
%

%  As usual, the \maketitle command creates the title and author/affiliations
%  display 

%\maketitle

%  If you are using the referee option, a new page, numbered page 1, will
%  start after the summary and keywords.  The page numbers thus count the
%  number of pages of your manuscript in the preferred submission style.
%  Remember, ``Normally, regular papers exceeding 25 pages and Reader Reaction 
%  papers exceeding 12 pages in (the preferred style) will be returned to 
%  the authors without review. The page limit includes acknowledgements, 
%  references, and appendices, but not tables and figures. The page count does 
%  not include the title page and abstract. A maximum of six (6) tables or 
%  figures combined is often required.''

%  You may now place the substance of your manuscript here.  Please use
%  the \section, \subsection, etc commands as described in the user guide.
%  Please use \label and \ref commands to cross-reference sections, equations,
%  tables, figures, etc.
%
%  Please DO NOT attempt to reformat the style of equation numbering!
%  For that matter, please do not attempt to redefine anything!

\section{Introduction}

The coronavirus disease (COVID-19) pandemic has shown profound impacts on public health and the economy worldwide. The development of efficient and effective public health interventions to prevent major outbreaks and contain the pandemic relies heavily on a quantitative understanding regarding the spread of the virus, such as the transmission rate and the average incubation period. A commonly used approach in epidemiology is to estimate these quantities of interest using epidemic mathematical models, such as the susceptible-infected recovered (SIR) model, with agent-based simulations which capture complex social networks and global scale into the models \citep{Spread,ModelingInfectious,Nature2009}. 

%how to use statistical approaches to model and understand the spread of COVID-19 to inform and educate the public about the virus transmission and develop effective strategies to contain the pandemic has become crucial.To develop public health policies to prevent major outbreaks and contain a pandemic
 
%{\bf Good reference: Modeling infectious disease dynamics in the complex landscape of global health, Science, 2015, Vol. 347, Issue 6227, aaa4339}

To estimate the parameters of interest, a widely used frequentist approach is to minimize the sum of the squared distances between the observed data and the simulation outputs, which is often referred to as the least squares approach. See, for example, \cite{chowell2003sars,chowell2004model,Capaldi2012, Chowell2017, PLOS2020,Bentout2020, Chen2020,Giordano2020}.  This estimation approach is intuitive and easy to compute; however, it is shown in this paper that this method is \textit{inefficient}, that is, its asymptotic variance is not theoretically minimal, particularly when the mathematical models associated with the simulators are built under certain assumptions
or simplifications, which may not hold in reality. These simulators are called \textit{imperfect} simulators in the computer experiment literature \citep{kennedy2001bayesian,tuo2015efficient,plumlee2017bayesian}. 
 Imperfect simulators are common in epidemiology \citep{ModelingInfectious}, and therefore estimate parameters of interest in epidemic models based on the least squares approach is not efficient.

To improve the estimation efficiency with imperfect epidemic models, a new estimation method is proposed in this paper. In the computer experiment literature, these unknown parameters associated with the mathematical models are often called \textit{calibration parameters}, and the process of estimating the parameters such that the model simulations agree with the observed data is called \textit{calibration} \citep{kennedy2001bayesian,santner2003design}. Although there are numerous developments on calibration, most of the work focus on continuous outputs while the discussions on non-Gaussian outputs, such as count data which are often observed in epidemiology, are scarce \citep{sung2017generalized,grosskopf2020generalized}. 
In this paper, we propose a new estimation method for non-Gaussian outputs, particularly for count data for our applications in epidemiology, which minimizes the $L_2$ projection of the discrepancy between the true mean process and the simulation outputs. It can be shown that the proposed estimator is asymptotically consistent, and provides a smaller asymptotic variance than the least squares estimator.
Furthermore, it can be shown that the proposed estimator achieves the semiparametric efficiency, even when the model simulations cannot match the reality due to certain assumptions or simplifications.

It is worth noting that there are extensive studies and applications of calibration by Bayesian procedures  \citep{Diekmann2013,farah2014bayesian,Song2020,Wulancet}. However, without taking the model imperfection into account in the conventional Bayesian framework, the theoretical justification for the parameter estimation with imperfect simulators are not fully developed. %estimating the  data-simulator discrepancy in the conventional Bayesian approach, theoretical justifications for the estimation properties with imperfect simulators are not well-developed.
On the other hand, Bayesian calibration of \cite{kennedy2001bayesian} takes into account the model imperfection through Gaussian process modeling, but it suffers from the \textit{unidentifiability issue} when the parameter estimation is of interest \citep{bayarri2007framework,han2009simultaneous,Hodges2010, Paciorek2010,gramacy2015calibrating}. Furthermore, most of the existing developments are based on continuous outputs with a Gaussian assumption, which is not valid for the count data in the epidemic models in our applications. Recent studies on addressing the unidentifiability issue can be found in \cite{plumlee2017bayesian} and \cite{tuo2017adjustments}.

The remainder of the paper is organized as follows. Two types of simulators for COVID-19 analysis, and a new estimation method based on $L_2$ projection for the unknown parameters in the simulators, are introduced in Section 2. Theoretical properties of the proposed estimator are developed in Section 3. In Section 4, numerical studies are conducted to demonstrate the finite sample performance of the proposed estimator and the empirical comparison with the least squares estimator. In section 5, the estimation method is applied to the study of COVID-19. Discussions and concluding remarks are given in Section 6. Computational details for the estimation are given in Appendix, and the mathematical proofs and the R \citep{R2018} code for implementation are provided in Supporting Web Materials.

\section{Estimation for Compartmental Models in Epidemiology}

\subsection{Imperfect Epidemic Models for COVID-19 Analysis}\label{sec:imperfectmodel}

Mathematical models are commonly used in epidemiology to provide scientific insights. These models are often developed based on certain simplifications of reality; therefore, they are imperfect \citep{ModelingInfectious}.
%Most of the mathematical models in epidemiology are imperfect \citep{ModelingInfectious}, therefore the parameters of interest estimated by minimizing the least squares distance are not efficient. 
For example, the SEIR model, which consists of four compartments, \textit{S}usceptible-\textit{E}xposed-\textit{I}nfectious-\textit{R}ecovered, is widely recommended for COVID-19 simulations because it accounts for the incubation period through the exposed compartment \citep{Wulancet,carcione2020simulation,mwalili2020seir,he2020seir,annas2020stability}, and is thus adopted in this paper.  
Mathematically, a deterministic SEIR model can be written as: 
\begin{equation}\label{eq:detSEIRmodel}
    \frac{dS}{dx}=-\frac{\beta IS}{N},\quad\frac{dE}{dx}=\frac{\beta IS}{N}-\kappa E,\quad\frac{dI}{dx}=\kappa E-\gamma I,\quad\frac{dR}{dx}=\gamma I,
\end{equation}
where $S$, $E$, $I$ and $R$ represent the numbers of cases in the corresponding compartment, $N=S+E+I+R$ is the total population, $x$ is time, $\beta$ is the contact rate that represents the average number of contacts per person per time in the susceptible compartment, $\gamma$ is the recovery rate from the infectious compartment, and $\kappa$ is the incubation rate which represents the rate of latent individuals becoming infectious, or equivalently, the average incubation period is $1/\kappa$. There are six unknown parameters  in the model \eqref{eq:detSEIRmodel}: $\beta,\kappa,\gamma$ and the initial numbers of infectious, exposed, and recovered cases (denoted by $I(0)$, $E(0)$, and $R(0)$ respectively), which are denoted by $\theta=(\beta,\kappa,\gamma,I(0),E(0),R(0))$.

In this paper, we focus on two types of SEIR simulators: a deterministic simulator and a stochastic simulator.
For a deterministic simulator, the simulation outputs are obtained by numerically solving the ordinary differential equations shown in (\ref{eq:detSEIRmodel}) using numerical solvers, such as the ODEPACK \citep{hindmarsh1983odepack}. 
On the other hand, a stochastic SEIR simulation provides a more sophisticated and realistic framework to integrate infection dynamics in different compartments as continuous-time Markov chains \citep{allen2008introduction,andersson2012stochastic,allen2017primer}. To conduct these simulations, we implement an \texttt{R} package, \texttt{SimInf}  \citep{SimInf2019}, in which the simulation results are obtained by the Gillespie stochastic algorithm \citep{Gillespie1977}. 
%Similar to the deterministic models, we consider 3 and 5 unknown parameters in the SIR and SEIR models, which are $(I_0,\beta,\gamma)$ and $(I_0,E_0,\beta,\kappa,\gamma)$, respectively.
Stochastic SEIR simulations are computationally more demanding. For example, it takes more than 10 minutes to produce one simulation result for one country under a given parameter setting. It is computationally infeasible to perform simulations for all the possible combinations of the parameters; therefore, an \textit{emulator} is constructed as an efficient surrogate to the actual simulation in our later implementation.

An accurate estimation of the unknown parameters in the SEIR model is often of great interest in epidemiology because it offers valuable insights into the dynamics of infectious diseases, which are essential for effectively predicting transmission patterns and assessing intervention strategies. For example, $1/\kappa$ indicates the average incubation period and 
the basic reproduction number, $R_0=\beta/\gamma$, represents the  expected number of new infected cases from an infectious individual in a population where all subjects are susceptible. An accurate and efficient estimation of these parameters is not only important for the public safety, but it also has significant impacts on global economy. 
The main objective in this paper is to provide a new estimation method that enhances the
estimation efficiency of parameters despite the inherent imperfections and limitations
of epidemic models.

\subsection{Least Squares Estimator and Maximum Likelihood Estimator}

%Compartmental models are widely used mathematical models for the study of infectious diseases, in which the population is assigned to compartments with labels such as \textit{S} (Susceptible), \textit{I} (Infectious), and \textit{R} (Recovered and immune) \citep{Diekmann2013}. For example, the most basic compartment model in epidemiology is the SIR model, in which each living individual is assumed to be in one of the above three compartments at any given time. The model specifies the rates at which individuals change their compartment, which include the progresses from $S$ to $I$ when infected, and from $I$ to $R$ upon recovery. Various extensions to SIR have been extensively developed in the literature of epidemiology \citep{ModelingInfectious}.

Let $f(x, \theta)$ denote the number of infected cases at time $x\in\Omega\subseteq\mathbb{R}^{+}$,  where $\theta\in\Theta\subseteq\mathbb{R}^q$ is a set of unknown calibration parameters associated with the compartmental model. %For instance, an SIR model contains two parameters, $\theta=(\beta,\gamma)$, where $\beta$ is the infection rate controlling the transition from $S$ to $I$, and $\gamma$ is the recovery rate controlling the transition from $I$ and $R$. 
In the case of SEIR model \eqref{eq:detSEIRmodel}, $q=6$ and $f(x, \theta)=\kappa E(x)$, where $E(x)$ is the solution of $E$ in the ordinary differential equations of \eqref{eq:detSEIRmodel}. Suppose that $y_i$ is the reported number of infected cases at time $x_i$. Then, given the reported number of infected cases in $n$ days, $\{(x_i,y_i)\}^n_{i=1}$, the commonly used approach to estimate the parameters is to minimize the sum of squared differences
between actual numbers of infected cases and simulation outputs from compartmental models. The estimated parameters are denoted by $\hat{\theta}_n^{\rm{LS}}$, where $LS$ stands for least squares, and they are obtained by
\begin{equation}\label{eq:ls_estimator}
    \hat{\theta}_n^{\rm{LS}}=\arg\min_{\theta\in\Theta}\sum^n_{i=1}(y_i-f(x_i,\theta))^2.
\end{equation} 

In addition to the least squares estimator, the maximum likelihood estimator (MLE) is also a commonly used estimation approach. Assume that $y_i\sim\text{Poi}(f(x_i,\theta))$, where $i=1,\ldots,n$, we obtain the MLE for the calibration parameters by
\begin{equation}\label{eq:mle}
\hat{\theta}_n^{\text{MLE}}=\arg\max_{\theta\in\Theta}\sum^n_{i=1}y_i\log f(x_i,\theta)-\sum^n_{i=1}f(x_i,\theta).
\end{equation}
 
\subsection{Estimate Calibration Parameters by $L_2$ Projection}\label{sec:estimation}

Despite the wide applications of the least squares approach and MLE, it can be shown that the least squares estimator does not achieve the semiparametric efficiency when the simulator $f(x,\theta)$ is imperfect, meaning that the simulation output cannot perfectly fit the response, even with the best fit of $\theta$. The asymptotic variance can be reduced by the proposed estimator introduced in this subsection. It can also be shown that MLE is asymptotically inconsistent when the simulator $f(x,\theta)$ is imperfect. Theoretical justifications are provided in Section \ref{sec:theoretical}.

Assume that the number of cases $y_i$ follows a Poisson distribution: $y_i\sim \text{Poi}(\lambda(x_i))$ for $i=1,\ldots, n$, and $y_i$ and $y_j$ are mutually independent for any $i\neq j$, where $\lambda(x_i)$ is the true mean function of $y_i$. The function $\lambda(x)$ is often called the \textit{true process} in the computer experiment literature \citep{kennedy2001bayesian,tuo2015efficient,tuo2016theoretical}. Ideally, if the underlying mean function $\lambda(x)$ is known, the true parameter can be defined as the minimizer of the $L_2$ projection of the discrepancy between the true process and the simulation output, that is,
\begin{equation}\label{eq:trueparameter}
    \theta^*=\arg\min_{\theta\in\Theta}\|\lambda(\cdot)-f(\cdot,\theta)\|_{L_2(\Omega)},
\end{equation}
where $\|g\|_{L_2(\Omega)}=\left(\int_{\Omega}g(x)^2\text{d}x\right)^{1/2}$.

%Ideally, if the underlying mean function $\lambda(x)$ is known, the true parameter can be defined as the minimizer of the expected squared discrepancy between the true process and the simulation output, that is,
%\begin{equation*}
%    \theta^*=\arg\min_{\theta\in\Theta}\mathbb{E}\left[\lambda(x)-f(x,\theta)\right]^2.
%\end{equation*}
%If we assume that $x$ is uniformly distributed, then it leads to the minimization of the $L_2$ projection of the discrepancy. That is, 
%\begin{equation}\label{eq:trueparameter}
%    \theta^*=\arg\min_{\theta\in\Theta}\|\lambda(\cdot)-f(\cdot,\theta)\|_{L_2(\Omega)},
%\end{equation}
%where \cmtC{$\|g\|_{L_2(\Omega)}=\left(\int_{\Omega}g(x)^2\text{d}x\right)^{1/2}$}.

In reality, the underlying true process $\lambda(\cdot)$ is unknown that needs to be estimated by observed data. Therefore, given the data $\{(x_i,y_i)\}^n_{i=1}$, we propose to estimate the true process by the kernel Poisson regression \citep{geer2000empirical,shim2011kernel}. Similar to the conventional Poisson regression \citep{mccullagh2019generalized}, we use the logarithm as the canonical link function, that is,  $\log\hat{\lambda}_n(\cdot)=\hat{\xi}_n(\cdot)$, and $\hat{\xi}_n(\cdot)$ is fitted by
\begin{equation}\label{eq:penalized_loglike}
    \hat{\xi}_n=\arg\min_{\xi\in\mathcal{N}_{\Phi}(\Omega)}\frac{1}{n}\sum^n_{i=1}\left(\exp\{\xi(x_i)\}-y_i\xi(x_i)\right)+\kappa_n\|\xi\|^2_{\mathcal{N}_{\Phi}(\Omega)},
\end{equation}
where $\|\cdot\|^2_{\mathcal{N}_{\Phi}(\Omega)}$ is the norm of the reproducing kernel Hilbert space generated by a given positive definite reproducing kernel $\Phi$, and $\kappa_n$ is a tuning parameter, which can be chosen by cross-validation methods. Thus, the proposed estimator of $\theta$, which we call \textit{$L_2$-estimator} throughout this paper, is the minimizer of the $L_2$ projection as follows:
\begin{equation}\label{eq:theta_hat}
    \hat{\theta}_n=\arg\min_{\theta\in\Theta}\|\hat{\lambda}_n(\cdot)-f(\cdot,\theta)\|_{L_2(\Omega)}.
\end{equation}
The optimal solution of \eqref{eq:penalized_loglike} has the form of $\hat{\xi}_n(x)=\hat{b}+\sum^n_{i=1}\hat{a}_i\Phi(x_i,x)$, where $\hat{b}$ and $\{\hat{a}_i\}^n_{i=1}$ can be obtained by the iterative re-weighted least squares algorithm \citep{green1985semi,hastie1990generalized,wahba1994soft}. The detail of the algorithm is given in Appendix \ref{append:algorithm}. In practice, the calculation of the $L_2$ norm in (\ref{eq:theta_hat}) can be approximated by numerical integration methods, such as Monte Carlo integration \citep{caflisch1998monte}.

%For some stochastic agent-based simulations in epidemiology that can be computationally intensive, 
As described in Section \ref{sec:imperfectmodel}, because stochastic SEIR simulations can be quite computationally intensive, it is infeasible to obtain $f(x,\theta)$ by conducting simulations for all possible combinations of the input parameters. Thus, we employ a computationally efficient \textit{emulator} to approximate the simulator. There are extensive studies on the development of statistical emulators in the computer experiment literature \citep{santner2003design}. 
Gaussian processes (GPs) are the most commonly used tools in the construction of emulators \citep{gramacy2020surrogates}.
Based on computer experiments with sample size $N$, a statistical emulator is denoted by $\hat{f}_N(x,\theta)$. which produces a predictive distribution of $f(x,\theta)$ with any untried $(x,\theta)\in(\Omega,\Theta)$. Specifically, the distribution of $\hat{f}_N(x,\theta)$ with any untried $(x,\theta)\in(\Omega,\Theta)$ is a normal distribution with the mean function, defined by $m_N(x,\theta)$ , and the variance function, defined by $v^2_N(x,\theta)$. We refer more details to \citep{gramacy2020surrogates}. Thus, by Fubini’s Theorem, the $L_2$-estimator of \eqref{eq:theta_hat} can be replaced by 
\begin{align}\label{eq:L2emulator}
    \tilde{\theta}_n&=\arg\min_{\theta\in\Theta}\mathbb{E}\|\hat{\lambda}_n(\cdot)-\hat{f}_N(\cdot,\theta)\|^2_{L_2(\Omega)}\nonumber\\
    &=\arg\min_{\theta\in\Theta}\int_{\Omega}\left(\hat{\lambda}_n(z)-m_N(z,\theta)\right)^2+v^2_N(z,\theta){\rm{d}}z.
\end{align}
%For these simulators, the proposed $L_2$-estimator can be similarly obtained by replacing  $f(x,\theta)$ in (\ref{eq:theta_hat}) with its emulator $\hat{f}(x,\theta)$. 
The applications of the proposed method with various existing emulators are demonstrated in Sections 4 and 5.

%For the time-series count data, the extension of \cite{sung2017generalized} which proposed an emulator for  time-series binary data. \cite{farah2014bayesian} proposed a dynamic model for agent-based models (write more here...). \cite{sung2020multiresolution} develop a computationally efficient emulator for large-scale computer simulations, which can be used for data which follows exponential-family distributions. \cite{binois2018practical} proposed a heteroscedastic Gaussian process and applied it to the emulation of a stochastic SIR model, which leverages the computational and statistical efficiency of designs under replication. 

%{\bf Revise: This estimator is chosen because of its asymptotic property, which is given in Supplementary Material \ref{sec:asym_klr}. The property is critical for the development of estimation consistency of $\hat{\theta}_n$ and its efficiency.}

It is worth noting that, the Poisson regression, $y_i\sim \text{Poi}(\lambda(x_i))$, may encounter \textit{overdispersion} due to the presence of greater variability \citep{mccullagh2019generalized}. That is, the variance of the data is larger than the mean, which violates the assumption of Poisson distribution. The deviance goodness of fit test \citep{mccullagh2019generalized} can be used to assess the model assumption. To take into account the issue of overdispersion, a \textit{quasi-Poisson} regression can be considered which assumes the variance of $y_i$ is $\phi\lambda(x)$, where $\phi>1$ is the \textit{overdispersion parameter}. The overdispersion parameter can be estimated by the ratio of the deviance to the effective degree freedom. The details of the deviance goodness of fit test and the estimation of overdispersion parameter are provided in Appendix \ref{append:algorithm}.

\section{Theoretical Properties}\label{sec:theoretical}

Theoretical properties of the $L_2$-estimator are discussed in this section, including the asymptotic consistency and the semiparametric efficiency. Theoretical comparisons with the least squares estimators are also provided by examining their asymptotic variances. The proofs are given in Supporting Web Materials.

The following theorem shows that the $L_2$-estimator $\hat{\theta}_n$ in (\ref{eq:theta_hat}) is asymptotically consistent and normally distributed.  

\begin{theorem}\label{thm:calibration}
Under the regularity conditions C1-C10 in Web Appendix B, we have
\begin{equation*}
    \sqrt{n}(\hat{\theta}_n-\theta^*)\xrightarrow{d}\mathcal{N}(0,4V_0(\theta^*)^{-1}W_0(\theta^*)V_0(\theta^*)^{-1}),
\end{equation*} 
as $n\rightarrow \infty$, where
\begin{equation}\label{eq:W0}
W_0(\theta)=\mathbb{E}\left[\lambda(X)\frac{\partial f}{\partial\theta}(X,\theta)\frac{\partial f}{\partial\theta^T}(X,\theta)\right]\quad{\rm{and}}\quad V_0(\theta)=\mathbb{E}\left[\frac{\partial^2}{\partial\theta\partial\theta^T}( \lambda(X)-f(X,\theta))^2\right].
\end{equation}
\end{theorem}

\begin{remark}
When the overdispersion parameter, $\phi$, is present in the Poisson regression, the result in Theorem \ref{thm:calibration} can be rewritten as 
\begin{equation*}
    \sqrt{n}(\hat{\theta}_n-\theta^*)\xrightarrow{d}\mathcal{N}(0,4\phi V_0(\theta^*)^{-1}W_0(\theta^*)V_0(\theta^*)^{-1}).
\end{equation*} 
\end{remark}

By the delta method, the following corollary  extends the result of Theorem \ref{thm:calibration} to a function of the $L_2$-estimator, which we denote as $g(\hat{\theta}_n)$.

\begin{corollary}\label{cor:deltamethod}
For a function $g$ satisfying the property that $\nabla g(\theta^*)$ exists and is  non-zero valued, we have 
\begin{equation*}
    \sqrt{n}(g(\hat{\theta}_n)-g(\theta^*))\xrightarrow{d}\mathcal{N}(0,4\nabla g(\theta^*)^TV_0(\theta^*)^{-1}W_0(\theta^*)V_0(\theta^*)^{-1}\nabla g(\theta^*)),
\end{equation*}
as $n\rightarrow \infty$.
\end{corollary}

Corollary \ref{cor:deltamethod} provides a theoretical support for the estimation and inference of some commonly used quantities of interest in epidemiology, such as the basic reproduction rate, which measures the transmission potential of a disease. For instance, in the SEIR model \eqref{eq:detSEIRmodel} the basic reproduction rate is a ratio of two of the calibration parameters, that is, $g(\theta)=\beta/\gamma$. The result of Corollary \ref{cor:deltamethod} can then be applied to construct the confidence intervals for the basic reproduction rate.

When estimating the unknown parameters in compartmental models, the parameter of interest $\theta^*$ in \eqref{eq:trueparameter} is $q$-dimensional, while the parameter space of the Poisson model, $y_i\sim \text{Poi}(\lambda(x_i))$, contains an infinite dimensional function space that covers $\lambda$.  Therefore, the calibration problem is regarded as a semiparametric problem. For these problems, the estimation method that can reach the highest estimation efficiency is called \textit{semiparametric efficient} \citep{bickel1993efficient,kosorok2008}. Specifically, let $\Lambda$ be an infinite dimensional parameter space whose true value is $\lambda_0$. Suppose that $T_n$ is an estimator for $\theta^*$ and $\sqrt{n}(T_n-\theta^*)$ is asymptotically normal. Let $\Lambda_0$ be an arbitrary finite dimensional space of $\Lambda$ that satisfies $\lambda_0\in\Lambda_0$. Consider the same calibration problem but with the parameter space $\Lambda_0$, then under this parametric assumption and some regularity conditions, an efficient estimator can be obtained by the maximum likelihood method, which is denoted by $S^{\Lambda_0}_n$. Then, the estimator $T_n$ is called \textit{semiparametric efficient} if there exists a $\Lambda_0$ such that $S^{\Lambda_0}_n$ has the same asymptotic variance as $T_n$. More details regarding the semiparametric efficiency can be found in \citep{tuo2015efficient,bickel1993efficient,kosorok2008}.
It can be shown in the following theorem that the proposed $L_2$-estimator is semiparametric efficient.

\begin{theorem}\label{thm:efficiency}
Under the regularity conditions in Theorem \ref{thm:calibration}, $\hat{\theta}_n$ is semiparametric efficient.
\end{theorem}

When the simulator $f$ is too costly to evaluate like the stochastic SEIR simulator in Section \ref{sec:imperfectmodel}, as discussed in Section \ref{sec:estimation}, an statistical emulator can be considered after conducting a computer experiment of size $N$ on the simulator. Suppose that the emulator of $f(x,\theta)$, i.e., $\hat{f}_N(x,\theta)$, follows a normal distribution with the mean function, $m_N(x,\theta)$, and the variance function, $v^2_N(x,\theta)$, i.e.,
\begin{equation}\label{eq:emulatorform}
    \hat{f}_N(x,\theta)\sim\mathcal{N}(m_N(x,\theta),v^2_N(x,\theta)),
\end{equation}
and the $L_2$-estimator is obtained by \eqref{eq:L2emulator} as $\tilde{\theta}_n$. Then, the following theorem provides the asymptotic distribution of $\tilde{\theta}_n$. 

\begin{theorem}\label{thm:thmwithemulator}
Under the regularity conditions C1 and C7-15 in Web Appendix B, we have
\begin{equation*}
    \sqrt{n}(\tilde{\theta}_n-\theta'_N)\xrightarrow{d}\mathcal{N}(0,4V_1(\theta'_N)^{-1}W_1(\theta'_N)V_1(\theta'_N)^{-1}),
\end{equation*} 
as $n\rightarrow \infty$, where
\begin{equation*}
   \theta'_N=\arg\min_{\theta\in\Theta}\|\lambda(\cdot)-m_N(\cdot,\theta)\|^2_{L_2(\Omega)}+\|\sqrt{v^2_N(\cdot,\theta)}\|_{L_2(\Omega)}^2
\end{equation*}

\begin{equation*}
W_1(\theta)=\mathbb{E}\left[\lambda(X)\frac{\partial m_N(X,\theta)}{\partial\theta}\frac{\partial m_N(X,\theta)}{\partial\theta^T}\right]\quad{\rm{and}}\quad V_1(\theta)=\mathbb{E}\left[\frac{\partial^2}{\partial\theta\partial\theta^T}\left(( \lambda(X)-m_N(X,\theta))^2+v^2_N(X,\theta)\right)\right].
\end{equation*}
\end{theorem}

With the emulator \eqref{eq:emulatorform}, it is of no surprise that the estimator $\tilde{\theta}_n$ is asymptotic inconsistent. However, when the size of the computer experiment, $N$, is sufficiently large, with an appropriate emulator (e.g., GP emulator) and under some regularity conditions, we have $m_N(x,\theta)\rightarrow f(x,\theta)$ and $v^2_N(x,\theta)\rightarrow 0$ for any $(x,\theta)\in\Omega\times\Theta$ \citep{wang2020prediction}, leading to $\theta'_N\rightarrow\theta^*$, which implies that $\tilde{\theta}_n$ is asymptotic inconsistent when $N$ is sufficiently large.

%\begin{remark}\label{rmk:emulator}
%Theorems \ref{thm:calibration} and \ref{thm:efficiency} remain valid even if the simulator $f$ is replaced by its emulator $\hat{f}$, provided that the emulator $\hat{f}$ satisfies the conditions C11-C12 given in Web Appendix B. 
%\end{remark}

In the next theorem, the asymptotic properties of the least squares estimator are developed and compared with those of the $L_2$-estimator.

\begin{theorem}\label{thm:OLS}
Under the regularity conditions C1-C4 and C16-C17 in Web Appendix B, we have 
\begin{equation*}
    \sqrt{n}(\hat{\theta}^{\rm{LS}}_n-\theta^*)\xrightarrow{d}\mathcal{N}(0,4V_0(\theta^*)^{-1}W_2(\theta^*) V_0(\theta^*)^{-1}),
\end{equation*} 
as $n\rightarrow\infty$, where
\begin{equation*}
    W_2(\theta)=W_0(\theta)+\mathbb{E}\left[(\lambda(X)-f(X,\theta))^2\frac{\partial f}{\partial\theta}(X,\theta)\frac{\partial f}{\partial\theta^T}(X,\theta)\right].
\end{equation*}
\end{theorem}

Similar to the $L_2$-estimator, it is shown that the least squares estimator is asymptotically consistent and  normally distributed. It can also be shown that $W_2(\theta^*)\geq W_0(\theta^*)$, which  leads to 
\begin{equation}\label{eq:inequalityTheorem5}
4V_0(\theta^*)^{-1}W_2(\theta^*)V_0(\theta^*)^{-1}\geq 4V_0(\theta^*)^{-1}W_0(\theta^*)V_0(\theta^*)^{-1}.
\end{equation}
This implies that the asymptotic variance of the least squares estimator $\hat{\theta}^{\rm{LS}}_n$ is greater  or equal to that of $\hat{\theta}_n$. The equality in \eqref{eq:inequalityTheorem5} holds if and only if 
\begin{equation}\label{eq:additionalterm}
    \mathbb{E}\left[(\lambda(X)-f(X,\theta^*))^2\frac{\partial f}{\partial\theta}(X,\theta^*)\frac{\partial f}{\partial\theta^T}(X,\theta^*)\right]=0.
\end{equation}
This result indicates that, if $\frac{\partial f}{\partial\theta}(x,\theta^*)\neq 0$ for all $x\in\Omega$, then \eqref{eq:additionalterm}  holds only if $\lambda(x)=f(x,\theta^*)$ for all $x\in\Omega$, which implies that the least squares estimator $\hat{\theta}^{\rm{LS}}_n$ is less efficient than $\hat{\theta}_n$ if $f$ is an imperfect simulator, i.e., $\lambda(x)\neq f(x,\theta^*)$ for some $x\in\Omega$.

In the next theorem, the asymptotic properties of the MLE as in \eqref{eq:mle} are developed and compared with those of the $L_2$-estimator.

\begin{theorem}\label{thm:MLE}
Under the regularity conditions C1 and C18-C22 in Web Appendix B, we have 
\begin{equation*}
    \sqrt{n}(\hat{\theta}^{\rm{MLE}}_n-\theta'')\xrightarrow{d}\mathcal{N}(0,V_3(\theta'')^{-1}W_3(\theta'') V_3(\theta'')^{-1}),
\end{equation*} 
as $n\rightarrow\infty$, where
\begin{equation}\label{eq:mleconverge}
    \theta''=\arg\max_{\theta\in\Theta}\mathbb{E}[\lambda(X)\log f(X,\theta)-f(X,\theta)],
\end{equation}
\begin{equation*}
    V_3(\theta)=\mathbb{E}\left[\frac{1}{f(X,\theta)}\left(-\frac{1}{2}V_0(\theta)+\left(1-\frac{\lambda(X)}{f(X,\theta)}\right)\frac{\partial f(X,\theta)}{\partial\theta}\frac{\partial f(X,\theta)}{\partial\theta^T}\right)\right],
\end{equation*}
and 
\begin{equation*}
    W_3(\theta)=\mathbb{E}\left[\frac{1}{f(X,\theta)^2}\left((\lambda(X)-f(X,\theta))^2+\lambda(X)\right)\frac{\partial f(X,\theta)}{\partial\theta} \frac{\partial f(X,\theta)}{\partial\theta^T} \right].
\end{equation*}
\end{theorem}

Unlike the $L_2$-estimator and the least squares estimator, the MLE asymptotically converges to a value that differs from the true parameter defined in \eqref{eq:trueparameter}. For example, suppose $\lambda(x)=x^2$ and $f(x,\theta)=\theta x$, by the definition of $\theta^*$ in \eqref{eq:trueparameter} we have $\theta^*=0.75$, while by \eqref{eq:mleconverge}, $\hat{\theta}_n^{\rm{MLE}}$ converges to $\theta''=2/3$ in probability.

\section{Numerical Study}

In this section, 
%we focus on numerical studies where the true parameters associated with the simulators have analytical solutions. These artificial examples are important to assess the performance of the proposed method because it is usually difficult to examine the performance in practice when the underlying true parameters are unknown.  
two artificial examples are conducted to examine the finite sample performance of the proposed method and compare the estimation performance with the least squares approach.
These numerical studies are conducted on a desktop with 3.5 GHz CPU and 8GB of RAM, and 4 CPUs are available for parallel computing.

\subsection{Imperfect simulator with one calibration parameter}\label{sec:1dexample}

\begin{figure}[h]
    \centering
    \includegraphics[width=\textwidth]{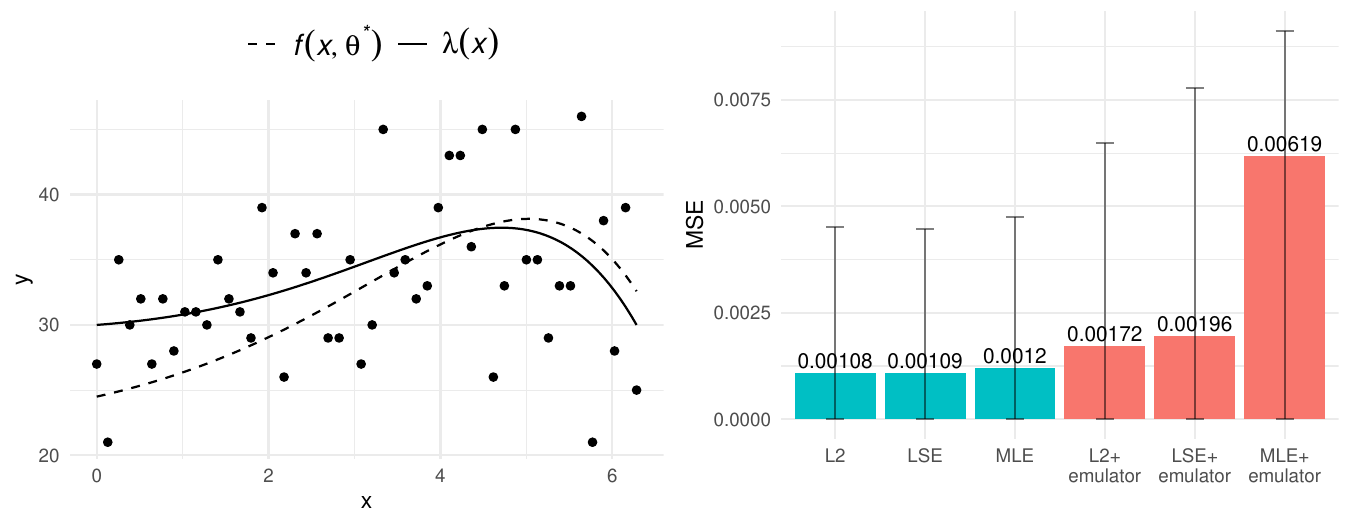}
    \caption{(Left) The true process $\lambda(x)$ as the solid line and the simulation output $f(x,\theta^*)$ as the dashed line. The real outputs are illustrated as the solid dots. (Right) Mean squared errors
    %, decomposed by bias squares (dark red) and variance (light blue), 
    of the estimates, where the error bars represent  the 5\% and 95\% quantiles.}
    \label{fig:example_1d}
\end{figure}

We consider an imperfect simulator adapted from \cite{tuo2015efficient} with one calibration parameter. The true process is assumed to be $\lambda(x)=\exp(x/2)\sin(x/2)+30$, where $x\in\Omega=[0,2\pi]$, and it is  illustrated in the left panel of Figure \ref{fig:example_1d} as the solid line. 
The data are generated from equal-spaced inputs in $[0,2\pi]$ with $n=50$ and the outputs are generated from a Poisson distribution with the mean process  $\lambda(x_i)$ for $i=1,\ldots,50$, which are shown as the solid dots in the left panel of Figure \ref{fig:example_1d}.

We assume that the simulation output is $f(x,\theta)=\lambda(x)-5\sqrt{\theta^2-\theta+1}(\sin(\theta x) + \cos(\theta  x))$, where  $\theta\in\Theta=[-1,1]$. This simulator is imperfect because $\sqrt{\theta^2-\theta+1}$ is always positive for any $\theta\in\Theta$. The true parameter can be analytically solved by minimizing \eqref{eq:trueparameter}, which gives that $\theta^*=-0.1789$.  
%The true process illustrated in the left panel of Figure \ref{fig:example_1d} by the solid line and the data $y_i$ generated from the true process are denoted by the solid dots. 
Plugging in the true calibration parameter, the simulator $f(x,\theta^*)$ is demonstrated as the dashed line, which is imperfect because, even with the true minimizer, the discrepancy between the simulation output and the true process is nonzero.

The performance of the $L_2$-estimator is compared with the least squares estimator and maximum likelihood estimator based on the mean squared errors (MSEs)  obtained from 100 replicates, that is, $\sum^{100}_{i=1}(\hat{\theta}_i-\theta^*)^2/100$, where $\hat{\theta}_i$ is the estimate at the $i$-th replicate. 
%Their MSEs are shown in the first three bars in the right panel of Figure \ref{fig:example_1d}, which are decomposed into squared biases (dark red) and variances (light blue). It shows that the $L_2$-estimator (``L2'') provides about $51\%$ smaller MSE than the least squares estimator (``LSE''). Such a reduction is mainly due to the reduction in the estimation variance of the $L_2$-estimator. On the other hand, the squared biases are relatively small for both estimators, which is consistent with the asymptotic result of estimation consistency in Section 3.  
Their MSEs are shown in the first three bars in the right panel of Figure \ref{fig:example_1d}. It shows that the $L_2$-estimator (``L2'') yields a smaller MSE than the least squares estimator (``LSE'') and maximum likelihood estimator (``MLE''). To quantify the uncertainty of the $L_2$ estimator, the  95\% confidence intervals are constructed based on the asymptotic result in Theorem \ref{thm:calibration}, where $\lambda,\theta^*$ and $\mathbb{E}$ are approximated by $\hat{\lambda}$, $\hat{\theta}_n$, and Monte-Carlo integration \citep{caflisch1998monte}, respectively. Out of the 100 replicates, the true parameter is contained by the confidence interval 96 times, which appears to be close to the nominal coverage 95\%.

We further compare the estimation performance for the cases when the simulations are computationally demanding and therefore statistical emulators are built as surrogates. Before comparing the estimation performance, we first examine the emulation performance of two existing emulation methods that are applicable to count data, which are the multiresolution functional ANOVA emulation \citep{sung2020multiresolution} and the heteroscedastic Gaussian process emulation \citep{binois2018practical}. Both methods have available packages in \texttt{R} \citep{R2018}, which are \texttt{MRFA} \citep{MRFA} and \texttt{hetGP} \citep{hetGP2019}, respectively. These emulators are trained by conducting a computer experiment, which simulates the model outputs of $f(x,\theta)$ of size $m$, where the inputs are sampled from $(x,\theta)\in(\Omega,\Theta)\in\mathbb{R}^2$ using a Latin hypercube design (LHD) \citep{mckay1979comparison}. For each input setting, simulations are conducted with $a$ replicates. The emulation performance is examined by performing predictions on $10,000$ random untried  input settings from $(\Omega,\Theta)$. With four different combinations of $m$ and $a$, the root mean squared prediction errors (RMSPEs) of the two emulators along with their computational time are reported in Table \ref{Tab:emulation_1d} of Appendix \ref{append:emulatorcomparison}. In this example, it appears that \texttt{hetGP} outperforms \texttt{MRFA} in terms of computational time and RMSPE. Thus, we select the emulator built by \texttt{hetGP} as the surrogate to the actual simulator in the following analysis.  %It is worth noting that in the \texttt{hetGP} method, the emulation accuracy can benefit more from adding more replicates while retaining the computational efficiency.

Next, we compare the estimation performance with the \texttt{hetGP} emulator built by $m=25, a=50$ samples, leading to total sample size $N=ma=750$. The $L_2$ estimator is obtained by \eqref{eq:L2emulator} with the emulator, and the least squares estimator is similarly obtained by minimizing $\sum^n_{i=1}(y_i-m_N(x_i,\theta))^2+v^2_N(z,\theta)$. For the MLE as in \eqref{eq:mle}, the actual simulator $f(x,\theta)$ is replaced by the mean of the \texttt{hetGP} emulator, i.e., $m_N(x,\theta)$. %both estimators, the actual simulator $f(x,\theta)$ is replaced by the \texttt{hetGP} emulator.
%The $L_2$-estimators are obtained by solving equation \eqref{eq:theta_hat} with the simulator replaced by the \texttt{hetGP} emulator. 
The MSEs are shown in the last three bars in the right panel of Figure \ref{fig:example_1d}. Similar to the previous result without emulators, the $L_2$-estimator provides a smaller MSE than the least squares estimator and MLE. By comparing the first three and last three bars, it is not surprising to see that the MSEs of ``{L2+emulator}'', ``{LSE+emulator}'', and ``{MLE+emulator}'' are larger than ``{L2}'', ``{LSE}'', and ``{MLE}'' due to the prediction uncertainty from emulation.  %It is also worth noting that the fourth bar is lower than the second and the third, which indicates that, by choosing the proposed estimator, the MSE can be smaller than the least squares estimator and MLE even if the simulations are approximated by an emulator. 
Similarly, we construct the 95\% confidence intervals based on the asymptotic result in Theorem \ref{thm:thmwithemulator} for the $L_2$ estimator of \eqref{eq:L2emulator}, and out of the 100 replicates, the true parameter is contained by the confidence interval 91 times, which appears to be close to the  nominal coverage of 95\%.

\iffalse
\begin{table}[ht]
    \centering
    \begin{tabular}{c|ccc}
    \toprule
     Method & Bias & SD & MSE\\
     \midrule
     Least squares    & 0.1282 & 0.3969& 0.1740\\
     New estimator   & 0.0376 & 0.2231& 0.0512\\
    \bottomrule
    \end{tabular}
    \caption{Caption}
    \label{tab:example_1d}
\end{table}
\fi

\subsection{Imperfect simulator with three calibration parameters}\label{sec:3dexample}

We consider a more complex problem with three calibration parameters adapted from \cite{plumlee2017bayesian}. Assume that the true mean process is $\lambda(x)=3x+3x\sin(5x)+3$ and the simulator is $f(x,\theta)=\theta_1+\theta_2x+\theta_3x^2$, where $x \in [0,2]$ and $\theta\in[0,5]^3$. Similar to the previous example, the three calibration parameters also have analytical solution $\theta^*\approx(3.56,0.56,1.76)$ by minimizing \eqref{eq:trueparameter}.

\begin{figure}[h]
    \centering 
    \includegraphics[width=0.95\textwidth]{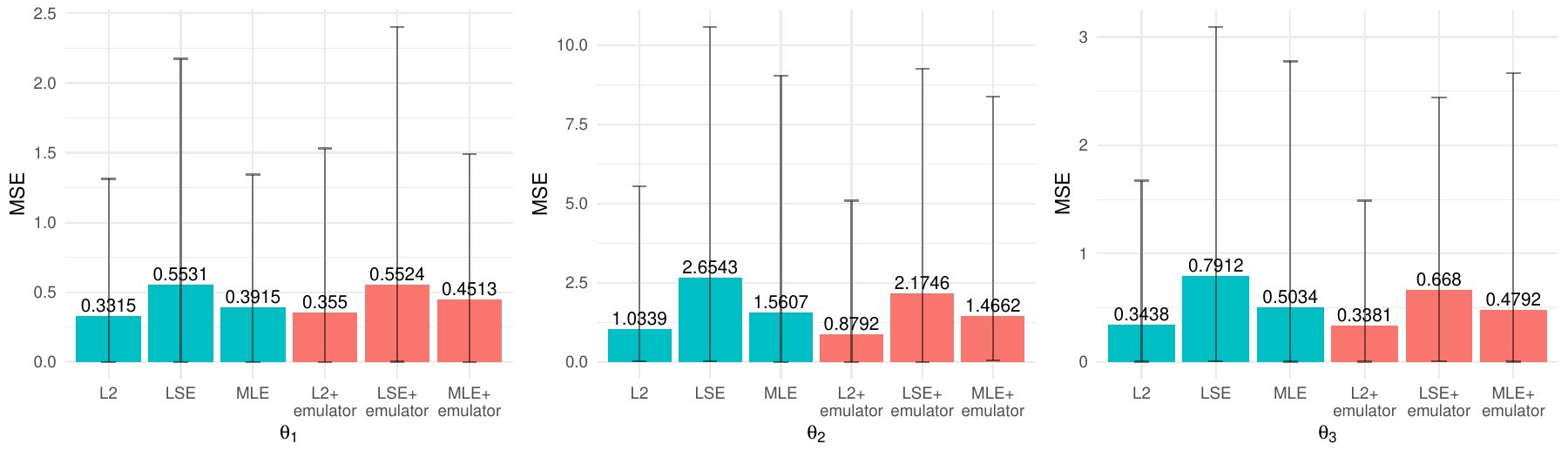}
    \caption{%Mean squared errors, decomposed by bias squares (dark red) and variances (light blue), of the estimates of (left) $\theta_1$, (middle) $\theta_2$, and (right) $\theta_3$
    Mean squared errors of the estimates of (left) $\theta_1$, (middle) $\theta_2$, and (right) $\theta_3$, where the error bars represent the 5\% and 95\% quantiles.}
    \label{fig:example_3d}
\end{figure}

The data $\{y_i\}^{50}_{i=1}$ are generated from the Poisson distribution with the mean $\{\lambda(x_i)\}^{50}_{i=1}$, where the 50 inputs are uniformly sampled from $[0,2]$. The estimation performance is examined based on the MSEs obtained from 100 replicates, and the proposed estimator and the least squares estimator are compared for each calibration parameter. The results are shown in the first three bars in each plot of Figure \ref{fig:example_3d}, in which the $y$-axis represents the MSEs. Similar to the previous example, %the MSEs are decomposed into squared biases (dark red) and the estimation variance (light blue). From these results, 
it appears that the $L_2$-estimator outperforms the least squares estimator and MLE for all of the three parameters.

In this example, we also examine the prediction performance of the two existing emulators, \texttt{MRFA} and \texttt{hetGP}. A computer experiment is conducted to train the two emulators by running the simulation outputs of $f(x,\theta)$ at $m$ unique sample locations with $a$ replicates, in which the unique input locations are sampled from $(x,\theta)\in(\chi,\Theta)\subseteq\mathbb{R}^4$ using an LHD. After the emulators are built, the RMSEs are computed based on the predictions of  $10,000$ untried input locations, and the prediction results are summarized in Table \ref{tab:emulation_3d} with different settings of $m$ and $a$. Similar to the previous example, the \texttt{hetGP} method outperforms \texttt{MRFA} in terms of prediction accuracy and computational time. With a larger $a$, i.e., more replicates, the prediction accuracy of \texttt{hetGP} appears to increase without much increase in computational time. Thus, we select \texttt{hetGP} as the emulator in the following analysis.

We now compare the estimation performance for the cases where emulators are constructed as surrogates to the actual simulations. The emulator is built by \texttt{hetGP} with $m=300,a=100$ and based on the emulator, the  estimation performance is summarized by the last three bars in each of the three plots in Figure \ref{fig:example_3d}. The results indicate that, either when the actual simulator is conducted or emulated, the $L_2$-estimator provides smaller MSEs compared to other two estimators.  

\iffalse
\begin{table}[!h]
\centering
\begin{tabular}{c|cccc}
\toprule
Method &  &  Bias & SD &MSE\\
\midrule
\multirow{3}{*}{Least squares} & $\theta_1$ &  -0.3935 & 0.7671 & 0.7433\\
&  $\theta_2$ &   1.2674 &1.8232 &4.9306\\
&  $\theta_3$ &   -0.7218 & 1.0392 & 1.6008\\
\midrule
\multirow{3}{*}{New estimator} &  $\theta_1$ &  -0.0316 & 0.6360 & 0.4054\\
&  $\theta_2$ &   0.1145& 1.3183 & 1.7511\\
  &  $\theta_3$ & -0.0811 & 0.8328 & 0.7001\\
\midrule
 &  $\theta_1$ & 0.0473 &  0.6191& 0.3855\\
New estimator &  $\theta_2$ & 0.0322  & 1.2011 & 1.4436\\
(with emulator)  &  $\theta_3$ & 0.3855& 0.9981 & 1.0321\\
\bottomrule
\end{tabular}
\caption{M}
\label{tbl:example_3d}
\end{table}
\fi

\section{Analysis of COVID-19}

We revisit the SEIR model in Section \ref{sec:imperfectmodel} and apply the proposed method to estimate the unknown parameters in the simulators for a better understanding of COVID-19 pandemic.  The estimation performance based on deterministic SEIR is discussed in Sections \ref{sec:realdata_deterministic} and the stochastic version is discussed in \ref{sec:realdata_stochastic}. 
To estimate the unknown parameters, we collect the actual numbers of infected cases from Johns Hopkins University CCSE repository \citep{dong2020interactive} through an \texttt{R} package \texttt{covid19.analytics} \citep{covid19package}. For each country, there are 365 observations collected from March 1st, 2020, to February 28th, 2021, denoted by $y_i$, where $i=1,\cdots, 366$. The studies are conducted for the top 20 countries which have the highest cumulative confirmed cases reported on March 1st, 2021.

\subsection{Parameter Estimation based on Deterministic SEIR}\label{sec:realdata_deterministic}
Before estimating the parameters, a deviance goodness of fit test is performed to examine the kernel Poisson regression as in \eqref{eq:penalized_loglike}, i.e., $y_i\sim \text{Poi}(\hat{\lambda}_n(x_i))$. It appears that the p-values of the test are all smaller than 0.0001, which indicates that there is a lack-of-fit in the current model. Therefore, a more flexible model, the quasi-Poisson as described in Section \ref{sec:estimation}, is applied to capture the potential overdispersion.

\begin{figure}[h]
    \centering
    \includegraphics[width=0.9\textwidth]{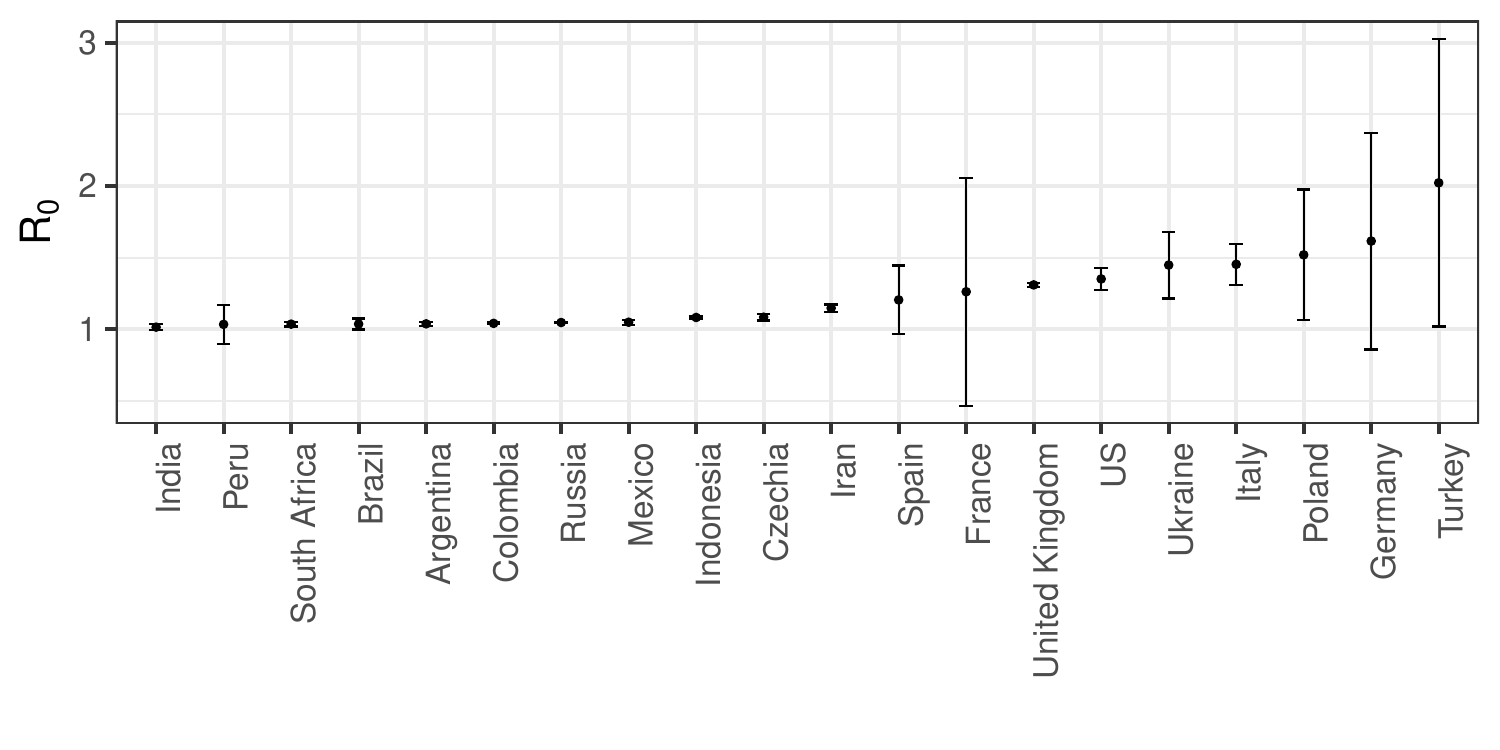}
    \caption{The estimated reproduction numbers for top-20 infectious countries based on the deterministic SEIR model.}
    \label{fig:deterministic_R0}
\end{figure}

For each country, the $L_2$-estimator of $\theta$ is obtained by minimizing \eqref{eq:theta_hat}, and the corresponding estimated reproduction number $R_0$ can be calculated by $R_0=\beta/\gamma$. The point estimates of $R_0$ and their $95\%$ confidence intervals, which are obtained by the result of Corollary \ref{cor:deltamethod}, are summarized in Figure \ref{fig:deterministic_R0} for the 20 countries. It shows that, from March, 2020 to March, 2021, all of the 20 countries have the basic reproduction numbers greater than 1, which means that the COVID-19 outbreak still post threats to these countries. Note that, the recovery rate $\gamma$ is in the denominator of $R_0$, and therefore the variation of $R_0$ appears to be higher for the countries having smaller recovery rates.
%only two countries have their basic reproduction number controlled below 1 and the COVID-19 outbreak still post threats to most of the countries.

\begin{figure}[h]
    \centering
    \includegraphics[width=0.9\textwidth]{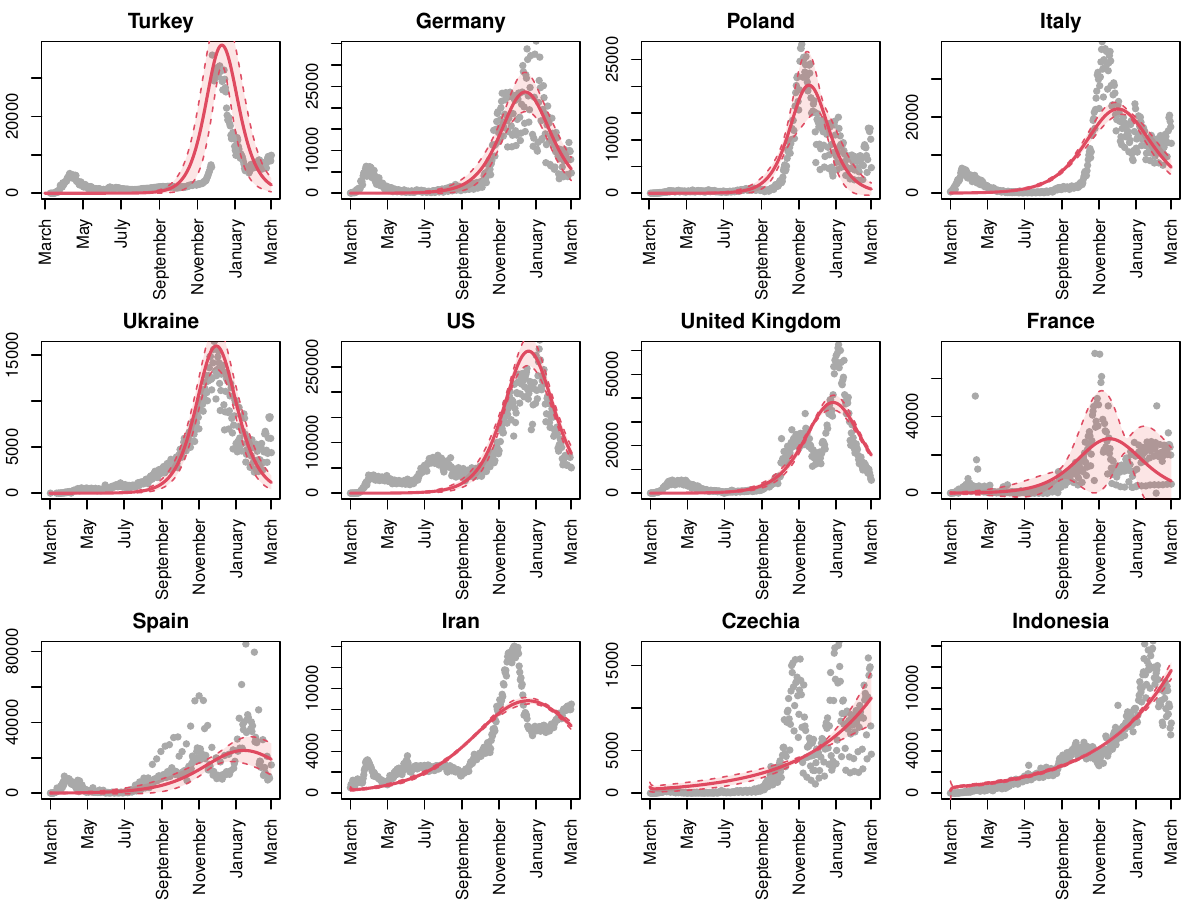}
    \caption{The gray dots are the actual numbers of daily infected cases. The red solid lines are the results from deterministic SEIR simulators by plugging in the $L_2$ estimates, and the red dashed lines are their corresponding 95\% confidence intervals.}
    \label{fig:deterministic_fit}
\end{figure}

Plugging in the $L_2$-estimators, the simulation results (solid lines), $f(x,\hat{\theta}_n)$, along with their confidence intervals (dashed lines), for the top 12 countries that have the highest $R_0$ values are demonstrated in Figure \ref{fig:deterministic_fit}. Note that the confidence intervals are similarly constructed based on Corollary \ref{cor:deltamethod}. That is, the variance of $f(x,\hat{\theta}_n)$ can be approximated by 
\begin{equation}\label{eq:detSEIRvar}
4\nabla_{\theta} f(x,\hat{\theta}_n)^TV_0(\hat{\theta}_n)^{-1}W_0(\hat{\theta}_n)V_0(\hat{\theta}_n)^{-1}\nabla_{\theta} f(x,\hat{\theta}_n),
\end{equation}
where $\nabla_{\theta}$ is the partial derivative with respect to $\theta$. In general, it appears that the simulation results can reasonably capture the overall trend observed from the actual numbers of infected cases, which are shown as the gray dots. For Iran, Czechia, and Spain, the discrepancy between the simulation results and actual observations is relatively larger than the other countries. This is partly because SEIR is an imperfect simulator which is built based on some assumptions or simplifications, and these assumptions may have larger deviations from the reality for certain countries. Another reason is that the intrinsic dynamics are  neglected in the deterministic simulations. To take into account the dynamics, a stochastic simulator is considered in the next subsection.
 
 %The fact that SEIR appears not to fit the data properly for COVID-19 was also pointed out in a recent study \citep{roda2020difficult}.

\subsection{Parameter Estimation based on Stochastic SEIR}\label{sec:realdata_stochastic}
 
%A stochastic SEIR simulation provides a more sophisticated and realistic framework to integrate infection dynamics in different compartments as continuous-time Markov chains \citep{allen2008introduction,andersson2012stochastic,allen2017primer}. To conduct these simulations, we implement an \texttt{R} package, \texttt{SimInf}  \citep{SimInf2019}, in which the simulation results are obtained by the Gillespie stochastic algorithm \citep{Gillespie1977}.  Stochastic SEIR simulations are computationally more demanding. For example, it takes more than 10 minutes to produce one simulation result for one country under a given parameter setting. It is computationally infeasible to perform simulations for all the possible combinations of the parameters; therefore, an emulator is constructed as an efficient surrogate to the actual simulation. 

Conducting stochastic simulations based on SEIR is computationally intensive, therefore emulators are developed as a faster surrogate to the actual stochastic simulations. In this study, we consider the \texttt{hetGP} emulator, which is built based on the simulations generated using a 60-run LHD for parameter settings with 20 equal-spaced time steps in $x$, which leads to the total sample size of $m=1200$. For each parameter-input setting, 50 replicates are simulated, i.e., $a=50$, so the total sample size of this computer experiment is $N=ma=60,000$. Based on this emulator, it takes less than two seconds to emulate the result for an untried parameter setting, which is significantly faster than the actual stochastic simulation.

\begin{figure}[!t]
    \centering
    \includegraphics[width=0.9\textwidth]{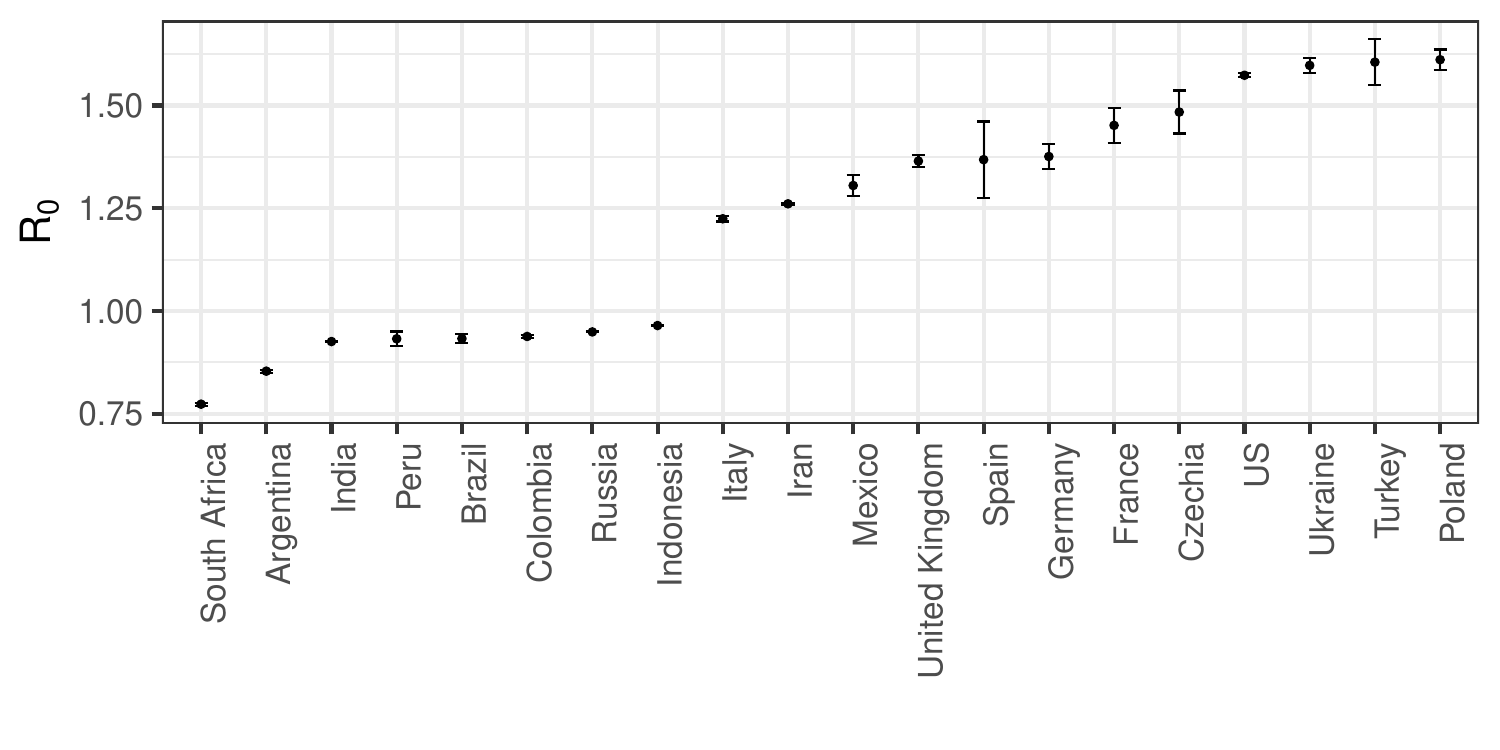}
    \caption{The reproduction numbers of top-20 infectious countries based on the stochastic SEIR model.}
    \label{fig:stochastic_R0}
        \centering
    \includegraphics[width=0.9\textwidth]{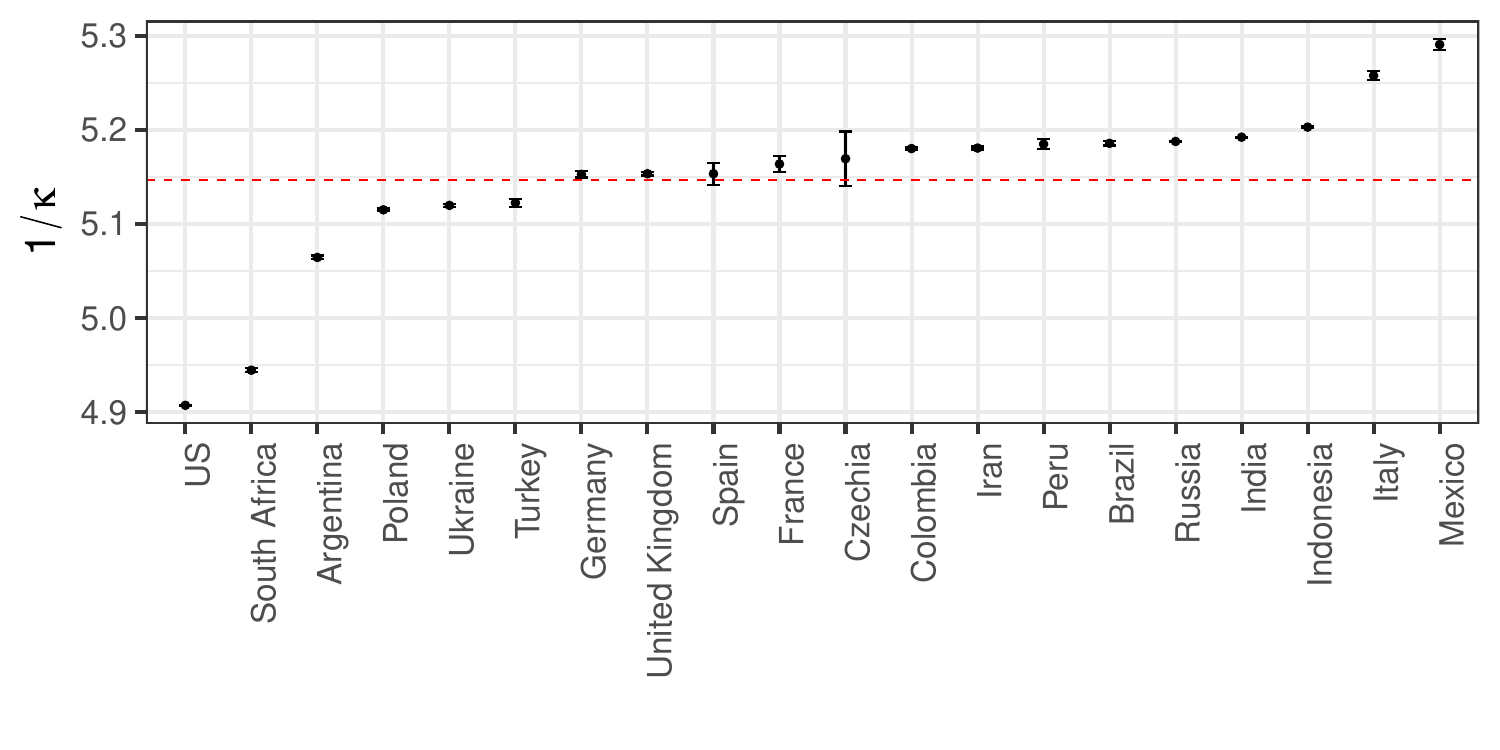}
    \caption{The estimated average incubation period based on the stochastic SEIR model, along with the overall average as indicated by the red dashed line.}
    \label{fig:stochastic_kappa}
\end{figure}

With the \texttt{hetGP} emulator, which has the form of  \eqref{eq:emulatorform}, the $L_2$-estimators are obtained by minimizing \eqref{eq:L2emulator}. The corresponding estimates of $R_0$ and their $95\%$ confidence intervals are summarized in  Figure \ref{fig:stochastic_R0}, where the variance is obtained based on the result of Theorem \ref{thm:thmwithemulator}. %The confidence intervals obtained from the stochastic simulations are wider than the deterministic ones because additional uncertainty from infection dynamics is incorporated. 
It appears that South Africa and Argentina  have their basic reproduction numbers controlled below 0.9,  which also show small basic reproduction numbers in the deterministic simulations (less than 1.05). We further report the estimated incubation period, $1/\kappa$, for each country and the corresponding $95\%$ confidence intervals in Figure \ref{fig:stochastic_kappa}. The overall average incubation period is 5.15 as indicated by the red dashed line. When comparing with the deterministic version, the estimation uncertainty based on the stochastic model is smaller. For example, the confidence intervals in Figure \ref{fig:stochastic_R0} are generally narrower than the ones in Figures \ref{fig:deterministic_R0}. The main reason is that the stochastic SEIR model accounts for the randomness and therefore the estimation is more robust to the noise, which leads to smaller uncertainty in the $R_0$ 
values compared to its deterministic counterpart. 
Having a slightly larger sample size for some countries may also be a factor of smaller uncertainty. Furthermore, we employed a frequentist framework and plugged the point estimate in the asymptotic variance in Corollary \ref{cor:deltamethod}, which may lead to an underestimation of the uncertainty from parameter estimation. To address this concern, an alternative approach is to adopt a Bayesian framework that incorporates prior distributions on the parameters. Further discussions regarding this Bayesian framework can be found in Section \ref{sec:discussion}.

%\begin{figure}[t]
%    \centering
%    \includegraphics[width=0.9\textwidth]{stochastic_kappa.pdf}
%    \caption{The estimated average incubation period based on the stochastic SEIR model.}
%    \label{fig:stochastic_kappa}
%\end{figure}

\begin{figure}
    \centering
    \includegraphics[width=0.9\textwidth]{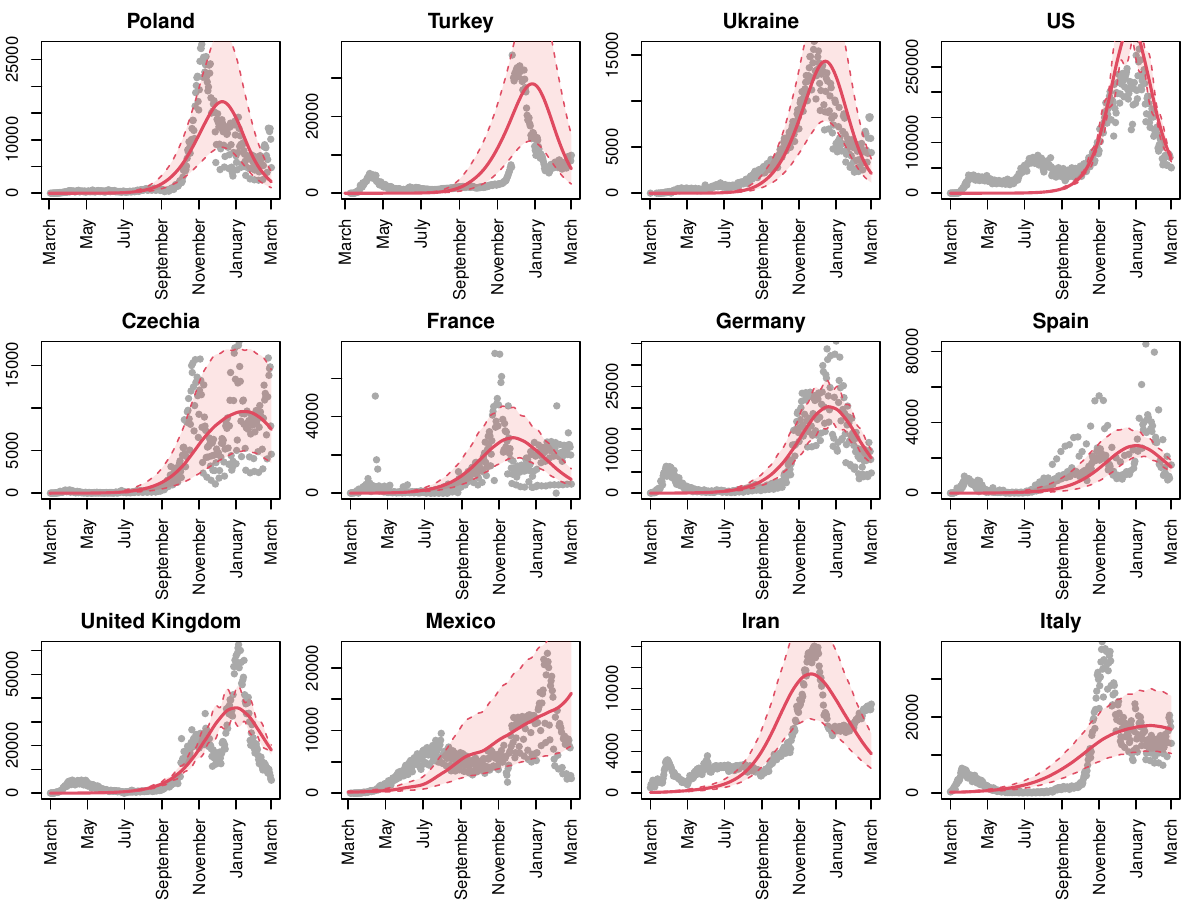}
    \caption{Number of infectious (gray dots) and the best fit of the stochastic SEIR models (red solid lines) of top-12 most infectious countries, where the red dashed lines are their corresponding 95\% confidence intervals.}
    \label{fig:stochastic_fit}
\end{figure}

In Figure \ref{fig:stochastic_fit}, the actual numbers of infected cases are illustrated as the gray dots. By plugging in the $L_2$-estimators, the simulation results for the top-12 countries with the highest $R_0$ are illustrated as the red curves, along with the $95\%$ confidence intervals as the red dashed lines. Overall, the simulation results show a much better agreement with the actual observations compared to the deterministic ones in Section \ref{sec:realdata_deterministic}. In particular, by taking into account the intrinsic dynamics, the simulation discrepancy for Czechia is significantly reduced from the deterministic one shown in Figure \ref{fig:deterministic_fit}. Note that the confidence intervals are computed based on $\mathbb{V}[\hat{f}_N(x,\tilde{\theta})]=\mathbb{E}[\mathbb{V}[\hat{f}_N(x,\tilde{\theta}_n)|\tilde{\theta}_n]]+\mathbb{V}[\mathbb{E}[\hat{f}_N(x,\tilde{\theta}_n)|\tilde{\theta}_n]]$, which can be approximated by 
\begin{equation}\label{stoSEIRvar}
v^2_N(x,\tilde{\theta}_n)+4\nabla_{\theta} m_N(x,\tilde{\theta}_n)^TV_1(\tilde{\theta}_n)^{-1}W_1(\tilde{\theta}_n)V_1(\tilde{\theta}_n)^{-1}\nabla_{\theta} m_N(x,\tilde{\theta}_n)
\end{equation}
using the result of Theorem \ref{thm:thmwithemulator}. 
When comparing with the predictive uncertainty of the deterministic model as shown in \eqref{eq:detSEIRvar}, the stochastic version as in \eqref{stoSEIRvar} introduces an additional source of uncertainty captured by the term $v^2_N(x,\tilde{\theta}_n)$, which accounts for the uncertainty due to emulation. This term contributes a dominating effect to the overall uncertainty, especially when stochastic models are computationally expensive and the emulators are constructed based on a limited number of computer experiments. 
%The first term, $v^2_N(x,\tilde{\theta}_n)$, is the emulation uncertainty uniquely presented in stochastic models due to its distinguishes the variance from that of the deterministic SEIR (see \eqref{eq:detSEIRvar}), which contains the errors sourced from the emulation uncertainty. 
As a result, even though the estimation uncertainty is relatively smaller with the stochastic model, the predictive uncertainty presented in Figure \ref{fig:stochastic_fit} is generally wider than the ones from the deterministic SEIR in Figure \ref{fig:deterministic_fit}.

\section{Discussions and Concluding Remarks}\label{sec:discussion}

Epidemic models for the analysis of COVID-19 are often imperfect. A new calibration method is proposed to estimate the unknown parameters in the imperfect epidemic models. The proposed estimator outperforms the least squares estimator by providing a smaller estimation variance and achieving the semiparametric efficiency. The proposed method is applied to the SEIR model for the analysis of COVID-19 pandemic. The estimates of the quantities of interest, such as the basic reproduction number and the average incubation period, and their confidence intervals are obtained based on the asymptotic results.

Apart from the frequentist approach studied in this paper, we are currently developing a Bayesian framework that extends the recent developments of Bayesian calibration to count data. For example, the orthogonal Gaussian process models  \citep{plumlee2016calibrating} or the Bayesian projected calibration \citep{tuo2017adjustments,xie2021bayesian} can be used to model the model discrepancy, which addresses the unidentifiability issue for continuous outputs, and it is conceivable to  further extend the modeling to count data by incorporating the idea of the generalized calibration in \cite{grosskopf2020generalized}. This framework is particularly useful when the goal is to provide a better fit to the data. Moreover, by incorporates prior distributions on the parameters and allowing for a range of plausible values, a Bayesian analysis can provide a more comprehensive assessment of uncertainty of the estimates. It is also worth investigating the confidence set on the calibration parameters using the method of \cite{plumlee2019computer} for the application herein. Another interesting direction that deserves further studies is to relax the constant parameter assumption. Instead, the calibration parameters can be assumed to be functions of some factors, such as time or temperature, which not only increases the model flexibility but also can provide further insights to the time-course dynamics of the COVID-19 infection. 
 
\section*{Acknowledgements} %The authors gratefully acknowledge helpful advice from the associate editor and two referees. 
This work was supported by NSF DMS 1660477 and NSF HDR TRIPODS award CCF 1934924.%\vspace*{-8pt}   

\section*{Supporting Web Materials}
Additional supporting information can be found online, including the mathematical proofs of Theorems \ref{thm:calibration}, \ref{thm:efficiency}, \ref{thm:thmwithemulator}, \ref{thm:OLS}, and \ref{thm:MLE}, and the R code for reproducing the results in the article.\vspace*{-8pt}

\section*{Appendix}
\begin{appendix}
\section{Algorithm to Estimate $\xi$ in (3) and Estimate overdispersion parameter $\phi$}\label{append:algorithm}
Since the optimal solution has the form of $\xi_n(x)=b+\sum^n_{i=1}a_i\Phi(x_i,x)$, one can show that the penalized likelihood in \eqref{eq:penalized_loglike} can be rewritten as $$
\frac{1}{n}\sum^n_{i=1}\left\{\exp\left(b+\mathbf{a}^T\boldsymbol{\psi}(x_i)\right)-y_i\left(b+\mathbf{a}^T\boldsymbol{\psi}(x_i)\right)\right\}+\kappa_n \mathbf{a}^T\boldsymbol{\Phi}\mathbf{a},
$$
where $\mathbf{a}=(a_1,\ldots,a_n)$, $\boldsymbol{\psi}(x)=(\Phi(x,x_1),\ldots,\Phi(x,x_n))$, and $\boldsymbol{\Phi}=(\Phi(x_i,x_j))_{1\leq i,j\leq n}$. The optimal solution of $\mathbf{a}$ and $b$ can then be obtained by taking the first-order partial derivatives of the objective function with respect to $\mathbf{a}$ and $b$ and setting them equal to zero, which can be solved by the 
iterative re-weighted least squares algorithm as follows. Denote
$$
\boldsymbol{\Phi}_0=\left(\begin{array}{cc} 0 & \mathbf{0}^T_n\\
\mathbf{0}_n & \boldsymbol{\Phi} \end{array}\right), \quad \boldsymbol{\Phi}_1=\left(\begin{array}{cc} \mathbf{1}_n & \boldsymbol{\Phi}\end{array}\right),
$$
where $\mathbf{1}_n=[1,\ldots,1]^T$ and $\mathbf{0}_n=[0,\ldots,0]^T$, and denote $\mathbf{W}$ as an $n\times n$ diagonal matrix with diagonal elements $\mathbf{W}_{ii}=\exp(b+\mathbf{a}^T\boldsymbol{\psi}(x_i))$. Then, in each step, one first solve for $\boldsymbol{\beta}:=(b,\mathbf{a}^T)^T$ in 
$$
\left(\boldsymbol{\Phi}_1^T\mathbf{W}\boldsymbol{\Phi}_1 +2n\kappa_n\boldsymbol{\Phi}_0\right)\boldsymbol{\beta}=\boldsymbol{\Phi}_1^T\mathbf{W}\boldsymbol{\eta},
$$
with an initial guess of $\boldsymbol{\eta}$, which is a vector of size $n$, and then update  each element of $\boldsymbol{\eta}$ by $$\boldsymbol{\eta}_i=(b+\mathbf{a}^T\boldsymbol{\psi}(x_i))+\frac{y_i-\exp(b+\mathbf{a}^T\boldsymbol{\psi}(x_i))}{\exp(b+\mathbf{a}^T\boldsymbol{\psi}(x_i))}.$$
The estimate $\hat{\boldsymbol{\beta}}$ can then be obtained by continuing solving for $\boldsymbol{\beta}$ and $\boldsymbol{\eta}$ iteratively until some convergence criterion is met.

To examine the goodness-of-fit of the Poisson regression, the following deviance goodness of fit test is considered. Since it can be shown that the deviance of the model follows a chi-square distribution asymptotically, that is
$$
D=2\sum^n_{i=1}\left(y_i\log(y_i/\hat{\lambda}_n(x_i))-(y_i-\hat{\lambda}_n(x_i))\right)\xrightarrow{d}\chi_{\text{edf}}
$$
when $n$ is sufficiently large, where the effective degree freedom, $\text{edf}=\text{trace}(\mathbf{S})$, where 
$$
\mathbf{S}=\boldsymbol{\Phi}_1\left(\boldsymbol{\Phi}_1^T\mathbf{W}\boldsymbol{\Phi}_1 +2n\kappa_n\boldsymbol{\Phi}_0\right)^{-1}\boldsymbol{\Phi}_1^T\mathbf{W}.$$
If the test indicates that overdispersion is present in the Poisson model, the overdispersion parameter $\phi$ can be estimated by 
$\hat{\phi}=D/\text{edf}$.

\section{Numerical Comparison of Emulators}\label{append:emulatorcomparison}

The numerical comparisons of the two emulators, \texttt{MRFA} and \texttt{hetGP}, for the numerical studies in Sections \ref{sec:1dexample} and \ref{sec:3dexample} are given in this section.

\begin{table}[!h]
\centering
\begin{tabular}{|C{2cm}|C{2cm}|C{2cm}|C{2cm}C{2cm}C{2cm}|}
\hiderowcolors
\hline \multirow{2}{*}{Emulator} & \multirow{2}{*}{$m$} & \multirow{2}{*}{$a$}  & Fitting  & Prediction & \multirow{2}{*}{RMSPE}\\
& &  & time (sec.) & time (sec.) & \\
\hline \multirow{4}{*}{\texttt{MRFA}} &  25 & 50 & 8 & 0.4 & 9.05\\
      & 25 & 100 & 11 & 0.4& 8.47\\
     & 50 & 50 & 11 & 0.7& 2.31\\
   & 100 & 100 & 29 & 0.7 & 0.99\\
\hline \multirow{4}{*}{\texttt{hetGP}} &  25 & 50  &  0.15 & 0.02 & 2.08\\
      & 25 & 100 & 0.15 & 0.02  & 1.74\\
     & 50 & 50 & 0.27 & 0.02 & 1.02\\
   & 100 & 100 & 1.16 & 0.07 & 0.50\\
\hline
\end{tabular}
\caption{Emulation performance for the example with one calibration parameter (in Section \ref{sec:1dexample}), where $m$ is the sample size of unique locations and $a$ is the number of replicates. RMSPEs are reported for the two emulators based on $10,000$ random predictive locations.}
\label{Tab:emulation_1d}
\end{table}

\begin{table}[t]
\centering
\begin{tabular}{|C{2cm}|C{2cm}|C{2cm}|C{2cm}C{2cm}C{2cm}|}
\hiderowcolors
\hline \multirow{2}{*}{Emulator} & \multirow{2}{*}{$m$} & \multirow{2}{*}{$a$}  & Fitting  & Prediction & \multirow{2}{*}{RMSPE}\\
& &  & time (sec.) & time (sec.) & \\
\hline \multirow{4}{*}{\texttt{MRFA}} &  300 & 50 & 258 & 3 & 0.66\\
      & 300 & 100 & 545 & 3& 0.63\\
     & 500 & 5 & 27 & 2& 0.82\\
   & 500 & 50 & 448 & 3 & 0.52\\
\hline \multirow{4}{*}{\texttt{hetGP}} &  300 & 50  &  7 & 1 & 0.20\\
      & 300 & 100 & 8 & 1  & 0.16\\
     & 500 & 5 & 29 & 2& 0.46\\
   & 500 & 50 & 29 & 2& 0.15\\
\hline
\end{tabular}
\caption{Emulation performance for the example with three calibration parameters (in Section \ref{sec:3dexample}), where $m$ is the sample size of unique locations and $a$ is the number of replicates. RMSPEs are reported for the two emulators based on $10,000$ random predictive locations.}
\label{tab:emulation_3d}
\end{table}

\end{appendix}

\bibliography{bib}

\begin{thebibliography}{60}
\providecommand{\natexlab}[1]{#1}
\providecommand{\url}[1]{\texttt{#1}}
\providecommand{\urlprefix}{}

\bibitem[{Funk et~al.(2009)Funk, S., and Gilad, E., and Watkins, C., and
  Jansen, V. A. A.}]{Spread}
Funk S, Gilad E, Watkins C, Jansen VAA.
\newblock The spread of awareness and its impact on epidemic outbreaks.
\newblock Proceedings of the National Academy of Sciences
  2009;106(16):6872--6877.

\bibitem[{Heesterbeek et~al.(2015)Heesterbeek, Hans and Anderson, Roy M and
  Andreasen, Viggo and Bansal, Shweta and De Angelis, Daniela and Dye, Chris
  and Eames, Ken TD and Edmunds, W John and Frost, Simon DW and Funk, Sebastian
  and Hollingsworth, T. Deirdre and House11, Thomas and Isham, Valerie and
  Klepac, Petra and Lessler, Justin and Lloyd-Smith, James O. and Metcalf, C.
  Jessica E. and Mollison, Denis and Pellis, Lorenzo and Pulliam, Juliet R. C.
  and Roberts, Mick G. and Viboud, Cecile and {Isaac Newton Institute IDD
  Collaboration}}]{ModelingInfectious}
Heesterbeek H, Anderson RM, Andreasen V, Bansal S, De~Angelis D, Dye C, et~al.
\newblock Modeling infectious disease dynamics in the complex landscape of
  global health.
\newblock Science 2015;347(6227):aaa4339.

\bibitem[{Epstein(2009)Joshua M. Epstein}]{Nature2009}
Epstein JM.
\newblock Modelling to contain pandemics.
\newblock Nature 2009;460(7256):687.

\bibitem[{Chowell et~al.(2003)Chowell, Gerardo and Fenimore, Paul W and
  Castillo-Garsow, Melissa A and Castillo-Chavez, Carlos}]{chowell2003sars}
Chowell G, Fenimore PW, Castillo-Garsow MA, Castillo-Chavez C.
\newblock {SARS} outbreaks in {Ontario, Hong Kong and Singapore}: the role of
  diagnosis and isolation as a control mechanism.
\newblock Journal of Theoretical Biology 2003;224(1):1--8.

\bibitem[{Chowell et~al.(2004)Chowell, Gerardo and Castillo-Chavez, Carlos and
  Fenimore, Paul W and Kribs-Zaleta, Christopher M and Arriola, Leon and Hyman,
  James M}]{chowell2004model}
Chowell G, Castillo-Chavez C, Fenimore PW, Kribs-Zaleta CM, Arriola L, Hyman
  JM.
\newblock Model parameters and outbreak control for SARS.
\newblock Emerging Infectious Diseases 2004;10(7):1258.

\bibitem[{Capaldi et~al.(2012)Capaldi, A., and Behrend, S., and Berman, B., and
  Simth, J., and Wright, J., and Lloyd, A. L.}]{Capaldi2012}
Capaldi A, Behrend S, Berman B, Simth J, Wright J, Lloyd AL.
\newblock Parameter estimation and uncertainty quantification for an epidemic
  model.
\newblock Mathematical Biosciences and Engineering 2012;9(3):553--576.

\bibitem[{Chowell(2017)Chowell, G.}]{Chowell2017}
Chowell G.
\newblock Fitting dynamic models to epidemic outbreaks with quantified
  uncertainty: A primer for parameter uncertainty, identifiability, and
  forecasts.
\newblock Infectious Disease Modelling 2017;2(3):379--398.

\bibitem[{Anastassopoulou et~al.(2020)Anastassopoulou, C., and Russo, L., and
  Tsakris, A., and Siettos, C.}]{PLOS2020}
Anastassopoulou C, Russo L, Tsakris A, Siettos C.
\newblock Data-based analysis, modelling and forecasting of the {COVID-19}
  outbreak.
\newblock PLoS One 2020;15(3):e0230405.

\bibitem[{Bentout et~al.(2020)Bentout, S., and Chekroun, A., and Kuniya,
  T.}]{Bentout2020}
Bentout S, Chekroun A, Kuniya T.
\newblock Parameter estimation and prediction for coronavirus disease outbreak
  2019 {(COVID-19)} in Algeria.
\newblock AIMS Public Health 2020;7(2):306--318.

\bibitem[{Chen and Qiu(2020)Chen, X., and Qiu, Z.}]{Chen2020}
Chen X, Qiu Z.
\newblock Scenario analysis of non-pharmaceutical interventions on global
  {COVID-19} transmissions.
\newblock Covid Economics: Vetted and Real-Time Papers, Centre for Economic
  Policy Research 2020;(7):46--67.

\bibitem[{Giordano et~al.(2020)Giordano, G., and Blanchini, F., and Bruno, R.,
  and Colaneri, P., Di Filippo, A., and Di Matteo, A., and Colaneri,
  M.}]{Giordano2020}
Giordano G, Blanchini F, Bruno R, Colaneri DFA P, Di~Matteo A, Colaneri M.
\newblock Modelling the {COVID-19} epidemic and implementation of
  population-wide interventions in {Italy}.
\newblock Nature Medicine 2020;26(6):855--860.

\bibitem[{Kennedy and O'Hagan(2001)Kennedy, Marc C and O'Hagan,
  Anthony}]{kennedy2001bayesian}
Kennedy MC, O'Hagan A.
\newblock Bayesian calibration of computer models.
\newblock Journal of the Royal Statistical Society: Series B
  2001;63(3):425--464.

\bibitem[{Tuo and Wu(2015)Tuo, Rui and Wu, C F Jeff}]{tuo2015efficient}
Tuo R, Wu CFJ.
\newblock Efficient calibration for imperfect computer models.
\newblock The Annals of Statistics 2015;43(6):2331--2352.

\bibitem[{Plumlee(2017)Plumlee, Matthew}]{plumlee2017bayesian}
Plumlee M.
\newblock Bayesian calibration of inexact computer models.
\newblock Journal of the American Statistical Association
  2017;112(519):1274--1285.

\bibitem[{Santner et~al.(2018)Santner, Thomas J and Williams, Brian J and Notz,
  William I}]{santner2003design}
Santner TJ, Williams BJ, Notz WI.
\newblock The Design and Analysis of Computer Experiments.
\newblock Second ed. Springer New York; 2018.

\bibitem[{Sung et~al.(2020)Sung, Chih-Li and Hung, Ying and Rittase, William
  and Zhu, Cheng and Wu, C F Jeff}]{sung2017generalized}
Sung CL, Hung Y, Rittase W, Zhu C, Wu CFJ.
\newblock A generalized {G}aussian process model for computer experiments with
  binary time series.
\newblock Journal of the American Statistical Association
  2020;115(530):945--956.

\bibitem[{Grosskopf et~al.(2020)Grosskopf, Michael and Bingham, Derek and
  Adams, Marvin L and Hawkins, W Daryl and Perez-Nunez,
  Delia}]{grosskopf2020generalized}
Grosskopf M, Bingham D, Adams ML, Hawkins WD, Perez-Nunez D.
\newblock Generalized Computer Model Calibration for Radiation Transport
  Simulation.
\newblock Technometrics 2020;in press.

\bibitem[{Diekmann et~al.(2013)Diekmann, O., and Heesterbeek, J. A. P., and
  Britton, T.}]{Diekmann2013}
Diekmann O, Heesterbeek JAP, Britton T.
\newblock Mathematical Tools for Understanding Infectious Disease Dynamics.
\newblock Princeton Univ. Press, Princeton; 2013.

\bibitem[{Farah et~al.(2014)Farah, Marian and Birrell, Paul and Conti, Stefano
  and Angelis, Daniela De}]{farah2014bayesian}
Farah M, Birrell P, Conti S, Angelis DD.
\newblock Bayesian emulation and calibration of a dynamic epidemic model for
  {A/H1N1} influenza.
\newblock Journal of the American Statistical Association
  2014;109(508):1398--1411.

\bibitem[{Wang et~al.(2020)Wang, L., and Zhou, Y., and He, J., and Wang, F.,
  and Tang, L., Eisenberg, M. and Song, P.}]{Song2020}
Wang L, Zhou Y, He J, Wang F, Tang EM L, Song P.
\newblock An epidemiological forecast model and software assessing
  interventions on {COVID-19} epidemic in China.
\newblock MedRxiv preprint 2020;.

\bibitem[{Wu et~al.(2020)Wu, J. T., and Leung, K., and Leung, G. M.}]{Wulancet}
Wu JT, Leung K, Leung GM.
\newblock Nowcasting and forecasting the potential domestic and international
  spread of the {2019-nCoV} outbreak originating in Wuhan, China: A modelling
  study.
\newblock The Lancet 2020;395(10225):689--697.

\bibitem[{Bayarri et~al.(2007)Bayarri, Maria J and Berger, James O and Paulo,
  Rui and Sacks, Jerry and Cafeo, John A and Cavendish, James and Lin, Chin-Hsu
  and Tu, Jian}]{bayarri2007framework}
Bayarri MJ, Berger JO, Paulo R, Sacks J, Cafeo JA, Cavendish J, et~al.
\newblock A framework for validation of computer models.
\newblock Technometrics 2007;49(2):138--154.

\bibitem[{Han et~al.(2009)Han, Gang and Santner, Thomas J and Rawlinson, Jeremy
  J}]{han2009simultaneous}
Han G, Santner TJ, Rawlinson JJ.
\newblock Simultaneous determination of tuning and calibration parameters for
  computer experiments.
\newblock Technometrics 2009;51(4):464--474.

\bibitem[{Hodges and Riech(2010)Hodges, J. S. and Riech, B. J.}]{Hodges2010}
Hodges JS, Riech BJ.
\newblock Adding spatially-correlated errors can mess up the fixed effect you
  love.
\newblock The American Statistician 2010;64(4):325--334.

\bibitem[{Paciorek(2010)Paciorek, C. J.}]{Paciorek2010}
Paciorek CJ.
\newblock The importance of scale for spatial-confounding bias and precision of
  spatial regression estimators.
\newblock Statistical Science 2010;25:107--125.

\bibitem[{Gramacy et~al.(2015)Gramacy, Robert B and Bingham, Derek and
  Holloway, James Paul and Grosskopf, Michael J and Kuranz, Carolyn C and
  Rutter, Erica and Trantham, Matt and Drake, R Paul and
  others}]{gramacy2015calibrating}
Gramacy RB, Bingham D, Holloway JP, Grosskopf MJ, Kuranz CC, Rutter E, et~al.
\newblock Calibrating a large computer experiment simulating radiative shock
  hydrodynamics.
\newblock The Annals of Applied Statistics 2015;9(3):1141--1168.

\bibitem[{Tuo(2019)Tuo, Rui}]{tuo2017adjustments}
Tuo R.
\newblock Adjustments to Computer Models via Projected Kernel Calibration.
\newblock SIAM/ASA Journal on Uncertainty Quantification 2019;7(2):553--578.

\bibitem[{{R Core Team}(2018)}]{R2018}
{R Core Team}.
\newblock R: A Language and Environment for Statistical Computing.
\newblock R Foundation for Statistical Computing, Vienna, Austria; 2018,
  \urlprefix\url{https://www.R-project.org/}.

\bibitem[{Carcione et~al.(2020)Carcione, Jos{\'e} M and Santos, Juan E and
  Bagaini, Claudio and Ba, Jing}]{carcione2020simulation}
Carcione JM, Santos JE, Bagaini C, Ba J.
\newblock A simulation of a {COVID-19} epidemic based on a deterministic {SEIR}
  model.
\newblock Frontiers in Public Health 2020;to appear.

\bibitem[{Mwalili et~al.(2020)Mwalili, Samuel and Kimanthi, Mark and Ojiambo,
  Viona and Gathungu, Duncan and Mbogo, Rachel Waema}]{mwalili2020seir}
Mwalili S, Kimanthi M, Ojiambo V, Gathungu D, Mbogo RW.
\newblock {SEIR} model for {COVID-19} dynamics incorporating the environment
  and social distancing.
\newblock BMC Research Notes 2020;to appear.

\bibitem[{He et~al.(2020)He, Shaobo and Peng, Yuexi and Sun,
  Kehui}]{he2020seir}
He S, Peng Y, Sun K.
\newblock {SEIR} modeling of the {COVID-19} and its dynamics.
\newblock Nonlinear Dynamics 2020;to appear.

\bibitem[{Annas et~al.(2020)Annas, Suwardi and Pratama, Muh Isbar and Rifandi,
  Muh and Sanusi, Wahidah and Side, Syafruddin}]{annas2020stability}
Annas S, Pratama MI, Rifandi M, Sanusi W, Side S.
\newblock Stability analysis and numerical simulation of {SEIR} model for
  pandemic {COVID-19} spread in {Indonesia}.
\newblock Chaos, Solitons \& Fractals 2020;to appear.

\bibitem[{Hindmarsh(1983)Hindmarsh, Alan C}]{hindmarsh1983odepack}
Hindmarsh AC.
\newblock {ODEPACK}, a systematized collection of ODE solvers.
\newblock Scientific Computing 1983;p. 55--64.

\bibitem[{Allen(2008)Allen, Linda J S}]{allen2008introduction}
Allen LJS.
\newblock An introduction to stochastic epidemic models.
\newblock In: Mathematical Epidemiology Springer; 2008.p. 81--130.

\bibitem[{Andersson and Britton(2012)Andersson, Hakan and Britton,
  Tom}]{andersson2012stochastic}
Andersson H, Britton T.
\newblock Stochastic Epidemic Models and Their Statistical Analysis.
\newblock Springer Science \& Business Media; 2012.

\bibitem[{Allen(2017)Allen, Linda J S}]{allen2017primer}
Allen LJS.
\newblock A primer on stochastic epidemic models: Formulation, numerical
  simulation, and analysis.
\newblock Infectious Disease Modelling 2017;2(2):128--142.

\bibitem[{Widgren et~al.(2019)Stefan Widgren and Pavol Bauer and Robin Eriksson
  and Stefan Engblom}]{SimInf2019}
Widgren S, Bauer P, Eriksson R, Engblom S.
\newblock {SimInf}: An {R} Package for Data-Driven Stochastic Disease Spread
  Simulations.
\newblock Journal of Statistical Software 2019;91(12):1--42.

\bibitem[{Gillespie(1977)Gillespie, D. T.}]{Gillespie1977}
Gillespie DT.
\newblock Exact Stochastic Simulation of Coupled Chemical Reactions.
\newblock The Journal of Physical Chemistry 1977;81(25):2340--2361.

\bibitem[{Tuo and Wu(2016)Tuo, Rui and Wu, C F Jeff}]{tuo2016theoretical}
Tuo R, Wu CFJ.
\newblock A theoretical framework for calibration in computer models:
  parametrization, estimation and convergence properties.
\newblock SIAM/ASA Journal on Uncertainty Quantification 2016;4(1):767--795.

\bibitem[{van~de Geer(2000)van de Geer, Sara}]{geer2000empirical}
van~de Geer S.
\newblock Empirical Processes in M-estimation.
\newblock Cambridge University Press; 2000.

\bibitem[{Shim and Hwang(2011)Shim, Jooyong and Hwang,
  Changha}]{shim2011kernel}
Shim J, Hwang C.
\newblock Kernel Poisson regression machine for stochastic claims reserving.
\newblock Journal of the Korean Statistical Society 2011;40(1):1--9.

\bibitem[{McCullagh and Nelder(2019)McCullagh, Peter and Nelder, John
  A}]{mccullagh2019generalized}
McCullagh P, Nelder JA.
\newblock Generalized linear models.
\newblock Second ed. New York: Routledge; 2019.

\bibitem[{Green and Yandell(1985)Green, Peter J and Yandell, Brian
  S}]{green1985semi}
Green PJ, Yandell BS.
\newblock Semi-parametric generalized linear models.
\newblock In: Proceedings 2nd International GLIM Conference, Lancaster, Lecture
  Notes in Statistics No. 32 New York: Springer; 1985.p. 44--55.

\bibitem[{Hastie and Tibshirani(1990)Hastie, Trevor and Tibshirani,
  Robert}]{hastie1990generalized}
Hastie T, Tibshirani R.
\newblock Generalized Additive Models.
\newblock New York: Chapman and Hall; 1990.

\bibitem[{Wahba et~al.(1995)Wahba, Grace and Gu, Chong and Wang, Yuedong and
  Campbell, R}]{wahba1994soft}
Wahba G, Gu C, Wang Y, Campbell R.
\newblock Soft classification, a.k.a. risk estimation, via penalized log
  likelihood and smoothing spline analysis of variance.
\newblock In: The Mathematics of Generalization, ed. D. H. Wolpert, Santa Fe
  Institute Studies in the Sciences of Complexity, Reading, MA: Addison-Wesley;
  1995. p. 329–--360.

\bibitem[{Caflisch(1998)Caflisch, Russel E.}]{caflisch1998monte}
Caflisch RE.
\newblock {M}onte {C}arlo and quasi-{M}onte {C}arlo methods.
\newblock Acta Numerica 1998;7(1):1--49.

\bibitem[{Gramacy(2020)Gramacy, Robert B}]{gramacy2020surrogates}
Gramacy RB.
\newblock Surrogates: Gaussian Process Modeling, Design, and Optimization for
  the Applied Sciences.
\newblock CRC Press; 2020.

\bibitem[{Bickel et~al.(1993)Bickel, Peter J and Klaassen, Chris A J and Ritov,
  Y and Wellner, Jon A}]{bickel1993efficient}
Bickel PJ, Klaassen CAJ, Ritov Y, Wellner JA.
\newblock Efficient and Adaptive Estimation for Semiparametric Models.
\newblock Johns Hopkins Univ. Press, Baltimore, MD.; 1993.

\bibitem[{Kosorok(2008)Kosorok, M. R.}]{kosorok2008}
Kosorok MR.
\newblock Introduction to Empirical Processes and Semiparametric Inference.
\newblock Springer, New York; 2008.

\bibitem[{Wang et~al.(2020)Wang, Wenjia and Tuo, Rui and Jeff Wu, C
  F}]{wang2020prediction}
Wang W, Tuo R, Jeff~Wu CF.
\newblock On prediction properties of kriging: Uniform error bounds and
  robustness.
\newblock Journal of the American Statistical Association
  2020;115(530):920--930.

\bibitem[{Sung et~al.(2020)Sung, Chih-Li and Wang, Wenjia and Plumlee, Matthew
  and Haaland, Benjamin}]{sung2020multiresolution}
Sung CL, Wang W, Plumlee M, Haaland B.
\newblock Multiresolution functional {ANOVA} for large-scale, many-input
  computer experiments.
\newblock Journal of the American Statistical Association
  2020;115(530):908--919.

\bibitem[{Binois et~al.(2018)Binois, Mickael and Gramacy, Robert B and
  Ludkovski, Mike}]{binois2018practical}
Binois M, Gramacy RB, Ludkovski M.
\newblock Practical heteroscedastic {G}aussian process modeling for large
  simulation experiments.
\newblock Journal of Computational and Graphical Statistics
  2018;27(4):808--821.

\bibitem[{Sung(2020)Chih-Li Sung}]{MRFA}
Sung CL.
\newblock MRFA: Fitting and Predicting Large-Scale Nonlinear Regression
  Problems using Multi-Resolution Functional ANOVA (MRFA) Approach; 2020, r
  package version 0.5.

\bibitem[{Binois and Gramacy(2019)Mickael Binois and Robert B.
  Gramacy}]{hetGP2019}
Binois M, Gramacy RB.
\newblock hetGP: Heteroskedastic Gaussian Process Modeling and Design under
  Replication; 2019, \urlprefix\url{https://CRAN.R-project.org/package=hetGP},
  r package version 1.1.1.

\bibitem[{McKay et~al.(1979)McKay, Michael D and Beckman, Richard J and
  Conover, William J}]{mckay1979comparison}
McKay MD, Beckman RJ, Conover WJ.
\newblock Comparison of three methods for selecting values of input variables
  in the analysis of output from a computer code.
\newblock Technometrics 1979;21(2):239--245.

\bibitem[{Dong et~al.(2020)Dong, Ensheng and Du, Hongru and Gardner,
  Lauren}]{dong2020interactive}
Dong E, Du H, Gardner L.
\newblock An interactive web-based dashboard to track {COVID-19} in real time.
\newblock The Lancet Infectious Diseases 2020;20(5):533--534.

\bibitem[{Ponce(2020)Marcelo Ponce}]{covid19package}
Ponce M.
\newblock covid19.analytics: Load and Analyze Live Data from the CoViD-19
  Pandemic; 2020,
  \urlprefix\url{https://CRAN.R-project.org/package=covid19.analytics}, r
  package version 1.1.

\bibitem[{Plumlee et~al.(2016)Plumlee, Matthew and Joseph, V Roshan and Yang,
  Hui}]{plumlee2016calibrating}
Plumlee M, Joseph VR, Yang H.
\newblock Calibrating functional parameters in the ion channel models of
  cardiac cells.
\newblock Journal of the American Statistical Association
  2016;111(514):500--509.

\bibitem[{Xie and Xu(2021)Xie, Fangzheng and Xu, Yanxun}]{xie2021bayesian}
Xie F, Xu Y.
\newblock Bayesian projected calibration of computer models.
\newblock Journal of the American Statistical Association
  2021;116(536):1965--1982.

\bibitem[{Plumlee(2019)Plumlee, Matthew}]{plumlee2019computer}
Plumlee M.
\newblock Computer model calibration with confidence and consistency.
\newblock Journal of the Royal Statistical Society: Series B
  2019;81(3):519--545.

\end{thebibliography}

\end{document}